\documentclass[twocolumn,showpacs,preprintnumbers,amsmath,amssymb]{revtex4}
\usepackage{graphicx}
\usepackage{dcolumn}

\newcommand{\lk}{\left( }
\newcommand{\rk}{\right)}
\newcommand{\ltk}{\left\{ }
\newcommand{\rtk}{ \right\} }
\newcommand{\ldk}{\left[ }
\newcommand{\rdk}{ \right] }

\begin{document}

\title{Baryons in Holographic QCD}
 

\author{%
Kanabu NAWA\footnote{E-mail: nawa@ruby.scphys.kyoto-u.ac.jp}, %
Hideo SUGANUMA\footnote{E-mail: suganuma@ruby.scphys.kyoto-u.ac.jp} %
and Toru KOJO\footnote{E-mail: torujj@ruby.scphys.kyoto-u.ac.jp}}

\affiliation{Department of Physics, Kyoto University, 
                Kyoto 606-8502, Japan}

\begin{abstract}
We study baryons 
in holographic QCD with
D4/D8/$\overline{\rm D8}$ multi D-brane system.
In holographic QCD, the baryon appears as a topologically non-trivial
chiral soliton 
in a four-dimensional effective theory of mesons. 
We call this topological soliton as Brane-induced Skyrmion.
Some review of D4/D8/$\overline{\rm D8}$ holographic QCD is presented from
the viewpoints
of recent hadron physics and QCD phenomenologies.
A four-dimensional effective theory with pions and $\rho$ mesons is
uniquely derived from the non-abelian Dirac-Born-Infeld (DBI) action
of D8 brane with D4 supergravity background
at the leading of large $N_c$, without small
amplitude 
expansion of meson fields to discuss chiral solitons.
For the hedgehog configuration of pion and $\rho$-meson fields,
we derive the energy functional and the Euler-Lagrange equation of
Brane-induced Skyrmion
from the meson effective action induced by holographic QCD.
Performing the numerical calculation, we obtain 
the soliton solution and figure out
the pion profile $F(r)$ and the $\rho$-meson profile $\tilde{G}(r)$  
of the Brane-induced Skyrmion with
its total energy, energy density distribution, and root-mean-square
radius.
These results are compared with the experimental quantities of baryons 
and also with the profiles of standard Skyrmion without $\rho$ mesons.
We analyze interaction terms of pions and $\rho$ mesons in
Brane-induced Skyrmion,
and find a significant $\rho$-meson component 
appearing in the core region of a baryon. 
\end{abstract} 
\maketitle
%
\section{Introduction}
%
Based on the recent remarkable progress in the concept
of gauge/gravity duality,
the non-perturbative properties of QCD, especially the low-energy
meson dynamics, are successfully described from 
the multi D-brane system consisting of D4/D8/$\overline{\rm D8}$ 
in the type IIA superstring theory~\cite{SS}.
This is called the Sakai-Sugimoto model, which is regarded as one of the reliable
holographic models of QCD.

In the D4/D8/$\overline{\rm D8}$ holographic QCD, 
the compositions of the four-dimensional massless QCD,
{\it i.e.}, 
gluons and massless quarks,
are represented in terms of the fluctuation modes of 
open strings on 
$N_c$-folded D4 branes
and
$N_f$-folded D8 and $\overline{\rm D8}$ branes.
%
Since the mass of D brane is proportional to the folding 
number of the D brane
as the Ramond-Ramond flux quantum,
the existence of the D brane with large folding number
can be described by a curved space-time.
Thus the supergravity description of D brane and also the 
construction of gauge theory from the D brane eventually 
gives the concept of `duality'
between the supergravity and gauge theory as the gauge/gravity duality
mediated by the mutual D brane.
This duality is firstly realized as the discovery of AdS/CFT
correspondence
by Maldacena in 1997~\cite{Mald}
as the duality between the type IIB superstring theory on ${\rm AdS}_5\times S^5$
and $N=4$ SUSY Yang-Mills theory through the D3
brane.
In the D4/D8/$\overline{\rm D8}$ holographic model,
with the large-$N_c$ condition as $N_f \ll N_c$, only D4 branes are 
represented by the classical supergravity background,
and back reaction from the probe D8 and $\overline{\rm D8}$ branes
to the total system is assumed to be small and neglected as 
a probe approximation.
With the existence of the heavy D4 brane, there appears a horizon in
ten-dimensional space-time as a classical D4 supergravity solution,
and probe D8 and $\overline{\rm D8}$ branes are interpolated with each
other in this curved space-time.
This interpolation induces
the spontaneous symmetry breaking of 
${\rm U}(N_f)_{\rm D8}\times {\rm U}(N_f)_{\overline{\rm D8}}$
local symmetry
into the single-valued ${\rm U}(N_f)_{\rm D8}$ symmetry,
which is regarded as the holographic manifestation of chiral symmetry
breaking.
The existence of D4 supergravity background is reflected on the metric 
of the non-abelian Dirac-Born-Infeld (DBI) action of D8 brane,
and after integrating out 
the symmetric directions along the
extra four-dimensional coordinate space,
the effective action of D8 brane is reduced into the 
five-dimensional Yang-Mills theory with flat four-dimensional space-time
$(t,{\bf x})$ and other extra fifth dimension with curved measure.
This five-dimensional Yang-Mills theory is regarded as a `unified' theory
of mesons,
including pions as massless pseudo-scalar Nambu-Goldstone bosons
with respect to chiral-symmetry breaking,
and infinite tower of massive (axial) vector mesons.
By diagonalizing this five-dimensional Yang-Mills action with 
regard to the parity eigen-states in the four-dimensional space-time,
meson spectrum is calculated, which is shown to achieve large
coincidence with experimental data for mesons.
Hidden local symmetry, which is phenomenologically introduced
to discuss the dynamics of pions and vector mesons~\cite{BKY},
naturally appears as a part of the local ${\rm U}(N_f)$ gauge symmetry
of D8 brane.
Other phenomenological hypothesis for the low-energy meson dynamics
like vector meson dominance~\cite{JJSa} and
Kawarabayashi-Suzuki-Riazuddin-Fayyazuddin relations among 
the couplings~\cite{KSRF}
are naturally reproduced in terms of the holographic QCD.

The D4/D8/$\overline{\rm D8}$ holographic model describes the two-flavor
three-color gauge theory as massless QCD from multi D-brane configurations.
However the reliability of the classical supergravity description of
D4 brane gives the constraint for the dual gauge theory side 
as a large 'tHooft coupling and also large $N_c$.
In this sense the holographic model describes the non-perturbative 
large-$N_c$ QCD.
According to the investigation about the large-$N_c$ gauge theory by
'tHooft in 1974~\cite{tHooft},
large-$N_c$ QCD is found to become equivalent with the weak coupling
system of mesons and glueballs; 
baryon does not directly appear as a dynamical degrees of freedom.
In 1979, Witten showed that, in the large-$N_c$ limit,
the mass of baryon increases and get proportional to $N_c$,
which explains the reason why baryon does not directly appear in the
large-$N_c$ limit~\cite{Witten}.
Witten also remarked that a baryon mass becomes proportional to
the inverse of the coupling constant of meson fields 
in the large $N_c$ limit,
similar to a mass of soliton wave in scalar field theories.
This viewpoint
gives the revival of the Skyrme model
describing a baryon as a chiral soliton.
%
From these considerations, a baryon is expected to appear as some
topological object like chiral soliton in the large-$N_c$ holographic QCD.

In this paper, 
we introduce the concept of chiral soliton
in the holographic QCD, {\it i.e.},
we describe a baryon as a topological soliton in the
four-dimensional meson effective action induced by the holographic QCD.
We call this topological soliton as Brane-induced Skyrmion.
Baryon as a mesonic soliton was firstly investigated without (axial)
vector mesons by Skyrme in 1961~\cite{Skyrme}, which is called Skyrme model,
basing on the non-linear sigma model as its meson effective action.
By comparing the properties of the standard Skyrmion without (axial)
vector mesons to Brane-induced Skyrmion,
the roles of (axial) vector mesons for baryons are discussed in our context.

This paper is organized as follows.
In section \ref{Mes}, we show the review of the D4/D8/$\overline{\rm D8}$ 
holographic model as one
of the reliable effective theories in holographic QCD.
There we also add new insights to this holographic model
from the viewpoints of recent hadron physics and QCD phenomenologies.
In the end of this section, four-dimensional meson effective action
is derived from the holographic QCD without small amplitude expansion
to discuss the chiral soliton as a large amplitude mesonic soliton.
We also show that the effects of higher mass excitation modes of (axial)
vector mesons for chiral solitons are expected to be small by introducing the
five-dimensional meson wave function picture.
Therefore we take only pions and $\rho$ mesons in the four-dimensional
meson effective action for the argument of chiral solitons in later
sections.
In section \ref{Bra},
we investigate Brane-induced Skyrmion by using the four-dimensional meson
effective action derived in the previous sections.
In subsection \ref{Sta}, we summarize the history of the standard 
Skyrmion with vector
mesons from the beginning of the Skyrme model in 1961.
The comparisons about the appearances of (axial) vector mesons
in the traditional phenomenological treatment like hidden local symmetry
approach~\cite{BKY} and those in Brane-induced Skyrmion as the
holographic QCD are discussed in detail.
In subsections \ref{Ans} and \ref{Ene},
configuration Ansatz as hedgehog type is taken for pion and $\rho$-meson
fields in the effective action,
and the energy function and Euler-Lagrange equation of Brane-induced
Skyrmion are derived.
In subsection \ref{Num},
numerical results about the profiles of Brane-induced Skyrmion are presented.
We find that stable soliton solution exists,
which should be a correspondent of a baryon in the large-$N_c$
holographic QCD.
Numerical results about the meson wave functions, energy density, total
energy, and root-mean-square radius of Brane-induced Skyrmion are
summarized,
where standard Skyrme profiles are also attached for comparison.
These results are compared with the experimental data for
baryons.
The effects of $\rho$ mesons for a baryon are also discussed 
from the holographic point of view.
Section \ref{Sum} is devoted to summary and discussion.
In Appendix~\ref{Ap}, topological natures of Skyrmions are summarized.
%
\section{Meson effective action from holographic QCD \label{Mes}}
%
In this section we review the D4/D8/$\overline{\rm D8}$ holographic model 
as one of the reliable effective theories in the holographic QCD~\cite{SS},
from the viewpoint of recent hadron physics and QCD phenomenologies. 
In subsection \ref{Eff}, following this holographic model,
we derive the effective action of pions and $\rho$ mesons
without small amplitude expansion of meson fields.
This action is used in later section \ref{Bra} to discuss
the chiral soliton, which is a non-trivial topological object
in the low-energy meson field theory.

\subsection{Holographic massless QCD with D4/D8/{\boldmath $\overline{\rm D8}$} \label{Hol}}
%
In this subsection we investigate the open-string spectrum of D4 and
D8-$\overline{\rm D8}$ branes in type IIA string theory,
from which we obtain  massless QCD in the weak coupling regime.

First, we review the construction of the four-dimensional pure Yang-Mills
theory from D4 brane configuration following Witten's strategy~\cite{WittenYM}.
In the superstring field theory,
D$p$ branes appear as $(p+1)$-dimensional
soliton-like objects in terms of the fundamental strings
in the ten-dimensional space-time.
In Witten's construction,
$N_c$-folded D4 branes as the supersymmetric five-dimensional manifolds 
are introduced extending to 
$x_{\mu=0\sim 3}$ and $x_4\equiv\tau$ directions in 
the ten-dimensional space-time.
From the fluctuation modes of 4-4 string 
($p$-$p^{'}$ string means an open string with one end located on D$p$ brane
and the other end on D${p^{'}}$ brane),
there exist ten massless bosonic modes
(${\cal A}_{\mu=0\sim 3}$, ${\cal A}_\tau$, $\Phi_{i=5\sim 9}$)
and massless fermionic modes as their superpartners.
The scalar modes $\Phi_{i=5\sim 9}$ are five Nambu-Goldstone modes
with respect to the spontaneous breaking of the translational invariance of D4 brane
in ten-dimensional space-time.

In order to obtain the non-SUSY Yang-Mills theory which includes only
massless gauge fields ${\cal A}_\mu$,
D4 brane is compactified to $S^1$ circle along the $\tau$ direction
with the periodicity of 
the Kaluza-Klein mass scale as $\tau \sim \tau+2\pi M_{\rm KK}^{-1}$,
and anti-periodic boundary condition in the $\tau$ direction is imposed
for all the fermions on D4 brane as
\begin{eqnarray}
\psi(x_\mu,\tau+2\pi M_{\rm KK}^{-1})=-\psi(x_\mu,\tau).\label{antiF1}
\end{eqnarray}
Then, fermion fields can be Fourier-transformed with boundary condition
(\ref{antiF1}) as
\begin{eqnarray}
\psi(x_\mu,\tau)\!&=&\!\!\sum_{n=-\infty}^{+\infty}\psi_n(x_\mu)%
                 e^{i\tau M_{\rm KK}\lk n+\frac{1}{2}\rk} \nonumber \\
 &\equiv&\!\! \sum_{n=-\infty}^{+\infty}\psi_n(x_\mu,\tau).\label{FourieF1}
\end{eqnarray}
These fermions are originally introduced as massless modes in five-dimensional space-time of
D4 brane. 
However the kinetic term of the fermion fields with the five-dimensional
Klein-Gordon operator
induces the large mass through the projection into the four-dimensional
space-time as
%
\begin{eqnarray}
\partial_M^2\psi_n \!&=&\!
(\partial_\mu^2 +\partial_\tau^2)\psi_n \nonumber \\
&=&\!\! \ltk\partial_\mu^2 -M_{\rm KK}^2\lk n+\frac{1}{2}\rk^2\rtk\psi_n, 
\nonumber \\
&& \hspace{3mm} (M=0\sim 4,\ \mu=0\sim 3)\label{KG1}
\end{eqnarray}
which means that, because of the boundary condition (\ref{antiF1}),
all the fermions acquire large masses
of Kaluza-Klein mass scale in four-dimensional space-time.
Then supersymmetry on D4 brane is completely broken, and
scalar fields ${\cal A}_\tau$ and $\Phi_{i=5\sim 9}$ are expected to acquire
large mass of order $O(M_{\rm KK})$ through loop corrections.
There also exists an infinite tower of massive Kaluza-Klein modes
originated from the 
compactified $S^1$ direction of D4
brane.
These extra modes for the Yang-Mills theory are 
expected to be neglected by taking large $M_{\rm KK}$ scale.
Then, only gauge fields ${\cal A}_{\mu=0\sim 3}$ are
left as massless modes and eventually give the 
pure Yang-Mills gauge theory.

Now, we discuss the construction of massless QCD in the weak coupling
regime by taking the multi D-brane configurations.
QCD is composed by two elements, {\it i.e.}, massless gauge fields (gluons) and light flavors
(quarks) as the matter fields.
In order to include the flavor degrees of freedom,
the idea of placing probe D6 brane in the D4 background is proposed
by Kruczenski {\it et al.}~\cite{KMMW}.
In their construction,
both D6 and D4 branes extend to the same flat four-dimensional
space-time $x_{\mu=0\sim 3}$ as their common part, and so
D6 and D4 branes do not extend to
all the directions of ten-dimensional space-time;
there remain two dimensions in which both D6 and D4 branes
can be seen as two localized points.
This means that the 4-6 string has generally finite length, and, 
because of its string tension,
it gives massive flavors as its fluctuation modes, 
which results in QCD with heavy flavors. 
In fact, D4/D6 model does not correspond to massless QCD, and, 
in low-energy regime, it does not have the spectrum of light pseudoscalar mesons
(pions) as the Nambu-Goldstone boson with respect to chiral-symmetry
breaking in QCD.

\begin{figure}[t]
  \begin{center}
   \resizebox{75mm}{!}{\includegraphics{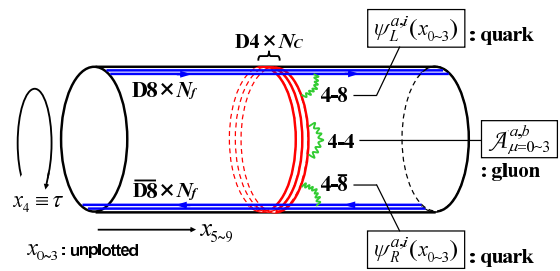}}
  \end{center}
\vspace{1.5mm}
\caption{ {\small 
Multi D-brane configurations of the D4/D8/$\overline{\rm D8}$
 holographic QCD. $N_c$-folded D4 branes and $N_f$-folded
D8-$\overline{\rm D8}$ branes. Flat four-dimensional space-time
$x_{0\sim 3}$ are not plotted. Gluons and quarks appear as the fluctuation modes of  
4-4, 4-8 and 4-$\overline{8}$ strings shown by the waving lines.} }
\label{D4D8D8_1_light}
\end{figure}

With these considerations, Sakai and Sugimoto~\cite{SS}
place the $N_f$-folded D8-$\overline{\rm D8}$ branes 
with the $N_c$-folded D4 branes as shown in 
Fig.\ref{D4D8D8_1_light} and Table~\ref{D4D8_conf}.
In this configuration, D8 and $\overline{\rm D8}$ branes are separately placed along the $\tau$
direction as shown in Fig.\ref{D4D8D8_1_light} to make tachyonic fluctuation modes of
$8$-$\overline{8}$ strings become massive
and negligible~\cite{SS}.
\begin{table}[floatfix]
\begin{center}
\begin{tabular}{ccccccccccc}
 {} & 0 & 1 & 2 & 3 & 4 & 5 & 6 & 7 & 8 & 9 \\ \hline
 D4 & $\bigcirc$ & $\bigcirc$ & $\bigcirc$ & $\bigcirc$ & $\bigcirc$ & %
    {} & {} & {} & {} & {}\\
 D8$\mbox{-}\overline{\rm D8}$ & $\bigcirc$ & $\bigcirc$ & $\bigcirc$ & $\bigcirc$ & %
    {} & $\bigcirc$ & $\bigcirc$ & $\bigcirc$ & $\bigcirc$ & $\bigcirc$\\ 
\end{tabular}
\caption{{\small Space-time extension of the D4 brane and
D8-$\overline{\rm D8}$ branes to construct massless QCD.  
The circle denotes the extended direction of each D-brane.
$x_{0 \sim 3}$ correspond to the flat four-dimensional space-time.
}}
\label{D4D8_conf}
\end{center}
\end{table}

Following Witten's construction mentioned above,  
D4 branes are compactified to $S^1$ circle in the extra direction $\tau(\equiv x_4)$ with 
the Kaluza-Klein mass scale $M_{\rm KK}$, 
and the anti-periodic boundary condition (\ref{antiF1}) is imposed
in the $\tau$ direction for all the fermions on D4 branes.
Owing to the compactification, only the massless gauge fields ${\cal A}_\mu$ are left,
and a non-SUSY four-dimensional ${\rm U}(N_c)$ gauge theory is obtained.
Here, the gauge fields ${\cal A}_\mu$ as the fluctuation modes of the 4-4 string 
belong to the adjoint representation of gauge group ${\rm U}(N_c)$,
and are identified as gluons in QCD.

Table~\ref{D4D8_conf} also shows that, comparing with D4/D6
model~\cite{KMMW},
D4 and D8-$\overline{\rm D8}$ branes extend to all the directions of the
ten-dimensional space-time.
This means that 4-8 string can be arbitrarily shorter,
especially localized around the physical-coordinate space
$x_{\mu=0\sim 3}$ shared by both D4 and D8-$\overline{\rm D8}$ branes
(see Table~\ref{D4D8_conf}).
Then 4-8 string gives 
$N_f$ flavors of massless fermions
as its fluctuation
modes localized around $x_{\mu=0\sim 3}$ in ten-dimensional space-time.
These flavors are shown to belong to the
fundamental representation of ${\rm U}(N_c)$ gauge group, and,
in this holographic model, they are interpreted as quarks in
QCD.

These flavors from 4-8 and 4-$\overline{8}$ strings are shown to have
opposite chirality with each other as the chiral fermions~\cite{ST}.
Therefore ${\rm U}(N_f)_{\rm D8}\times {\rm U}(N_f)_{\overline{\rm D8}}$
local gauge symmetry of probe D8-$\overline{\rm D8}$ branes
can be interpreted as the correspondent of
${\rm U}(N_f)_L\times{\rm U}(N_f)_R$ chiral
symmetry
for flavor quarks in QCD.
These considerations show that
the D4/D8/$\overline{\rm D8}$ holographic model
represents the four-dimensional massless QCD.

\subsection{Supergravity description of background D4 brane \label{Sup}}
In the previous section, we showed that
massless QCD can be constructed from the
D4/D8/$\overline{\rm D8}$ multi-brane configurations
by considering the symmetry of the induced gauge theory
with the analysis of the open-string spectrum.
On the other hand, 
according to the recent remarkable progress in the concept of
gauge/gravity duality~\cite{AGMOO},
D brane has a supergravity description, and,
through this duality, the parameters among the gauge theory and the string theory 
(and also the supergravity solution)
are found to be related with each other.
In the actual calculations, classical supergravity solutions are
tractable, and 
the reliability of this tree-level approximation with classical supergravity
solution gives some constraints for the parameters ($g_{\rm YM}$, $N_c$) 
of the dual gauge theory.
In this section, we show the relations between the gauge theory and 
the supergravity description of D branes, 
and discuss the constrained properties of massless QCD
which is taken to be dual with classical supergravity description
of D branes.

Here, we apply the probe approximation for the D4/D8/$\overline{\rm D8}$ system;
$N_f$-folded D8-$\overline{\rm D8}$ flavor branes are introduced as the probes
into the supergravity background solution of the 
$N_c$-folded D4 branes,
and back reaction from the probe branes to the total system is neglected.
This D4 background solution is shown to be holographic dual with
the four-dimensional pure Yang-Mills theory at low energies~\cite{WittenYM}.
This probe approximation is thought to be reliable 
for the large-$N_c$ case as $N_f \ll N_c$,
since the folding number of D brane, {\it i.e.},
the Ramond-Ramond flux quantum,
is proportional to the mass of D brane.
The probe approximation is sometimes
compared to the general coordinate system with
a light probe object and the gravitational background induced by a
heavier object.

This probe approximation is somehow similar to the quenched approximation, 
which is frequently used in lattice QCD calculations~\cite{R05}.
In quenched QCD, 
corresponding to $N_f=0$ limit,
the nonperturbative QCD vacuum is composed only with the gluonic part of QCD, 
and the (anti)quarks are treated as `probe' fermions traveling in the gluonic vacuum.
In fact, the quark-loop effect is neglected in the quenched approximation like large $N_c$ QCD,
and the `back reaction' of the quark field to the nonperturbative gluonic vacuum is neglected. 
In spite of such reduction of quark dynamics, 
the quenched approximation with $N_c=3$ is known to reproduce 
the low-lying hadron spectra only within 10\% deviation 
in lattice QCD calculations
relative to the experimental data,
as well as 
the nonperturbative aspects of QCD such as the quark confinement properties 
in the inter-quark potentials and dynamical chiral-symmetry breaking~\cite{R05}.
These phenomenological success of 
the quenched approximation in QCD
may be related with the viewpoint of the probe approximation in the
holographic approach of QCD.

The supergravity solution of $N_c$-folded D4 branes compactified along the
$\tau$ direction to $S^1$ circle can be written 
in the Euclidean metric as follows:
\begin{eqnarray}
\hspace*{4mm}ds^2 \!&=&\! \lk\frac{U}{R}\rk^{3/2}\lk%
      g_{\mu\nu}dx_{\mu}dx_{\nu}+f\lk U\rk d\tau^2\rk \nonumber \\
     && + \lk\frac{R}{U}\rk^{3/2}\lk%
     \frac{dU^2}{f\lk U\rk}+U^2 d\Omega_4^2\rk, \nonumber\\ 
e^{\phi} \!&=&\! g_s\lk\frac{U}{R}\rk^{3/4},          \nonumber\\ 
F_4 \!&=&\! dC_3=\frac{2\pi N_c}{V_4}\epsilon_4,      \nonumber\\ 
f(U) \!&=&\! 1-\frac{U_{\rm KK}^3}{U^3}.              \label{D4_sugra}
\end{eqnarray}
$ds$ is the infinitesimal invariant distance
in ten-dimensional space-time: 
$d s^2\equiv g_{MN} dx_M dx_N $ ($M,N=0\sim 9$).
$x_{\mu=0\sim 3}$ and $x_4\equiv \tau$ are the coordinates to which
D4 brane extends, and
$x_{\mu=0\sim 3}$ have the flat four-dimensional Euclidean metric
$g_{\mu\nu}$.
D4 brane does not extend to the residual five dimensions
$x_{\alpha=5\sim 9}$, so that its classical supergravity solution has 
${\rm SO}(5)$ rotational symmetry around $x_{\alpha=5\sim 9}$,
which is usefully taken as the polar coordinate $(U,\Omega_4)$;
$U$ corresponds to the radial coordinate and $\Omega_4$ are four angle
variables in $x_{\alpha=5\sim 9}$ directions, 
parametrizing a unit sphere $S^4$.
$\phi$ is a dilaton field from the fluctuation modes of closed strings.
$F_4$ is the 4-form field strength and $C_3$ is the 3-form Ramond-Ramond field.
$\epsilon_4$ is a volume form on $S^4$ with total volume $V_4=8\pi^2/3$.
The radial coordinate $U$ is bounded from below as
$U\ge U_{\rm KK}$, and, at $U=U_{\rm KK}$, the $S^1$ circle around $\tau$
direction completely shrinks with $f(U)\rightarrow 0$;
$U_{\rm KK}$ corresponds to a horizon in ten-dimensional space-time
with the D4 supergravity solution.
The constant $R$ can be written by the string coupling constant
$g_s\equiv e^{<\phi>}$
and string length $l_s$~\cite{SS} as
\begin{eqnarray}
R^3=\pi g_s N_c l_s^3. \label{R_def}
\end{eqnarray}

To avoid a conical singularity at $U=U_{\rm KK}$ and make the solution
(\ref{D4_sugra}) everywhere regular,
the period $\delta\tau$ of the compactified $\tau$ direction is taken~\cite{SS} to be 
\begin{eqnarray}
\delta\tau\equiv \frac{4\pi}{3}\frac{R^{3/2}}{U_{\rm KK}^{1/2}}.\label{delta_tau1}
\end{eqnarray}

As discussed in the previous section about the open-string spectrum on
the D4 brane,
the four-dimensional pure Yang-Mills theory becomes effectively the same as
the gauge theory induced on the compactified D4 brane below the 
Kaluza-Klein mass scale $M_{\rm KK}$,
which is related to the period $\delta\tau$ as
\begin{eqnarray}
M_{\rm KK}=\frac{2\pi}{\delta\tau}=\frac{3}{2}\frac{U_{\rm KK}^{1/2}}{R^{3/2}}.\label{M_KK_deltatau}
\end{eqnarray}

The Yang-Mills coupling constant $g_{\rm YM}$ of the gauge theory
can be also written by the parameters of the string theory
by using the DBI action of D4 brane as
\begin{eqnarray}
S_{\rm D4}^{\rm DBI} \!&=&\! T_4\int d^4 x d\tau
               e^{-\phi}\sqrt{-\det\lk g_{MN}+2\pi\alpha^{'}{\cal F}_{MN}\rk}
	       \nonumber\\
            &\sim &\!  T_4\frac{1}{4}(2\pi\alpha^{'})^2 \frac{\delta\tau}{g_s}
                   \int d^4 x 2\mbox{tr} {\cal F}_{\mu\nu}^2 +O({\cal F}^4),
		   \nonumber \\
  &&\hspace{8mm}\mbox{($M,N=0\sim 4$,\hspace{2mm} $\mu,\nu=0\sim 3$)}\label{DBI_D4}
\end{eqnarray}
where $\alpha^{'}\equiv l_s^2$ is the Regge slope parameter and
$T_p\equiv \frac{1}{\lk 2\pi\rk^p l_s^{p+1}}$ is the surface tension
of D$p$ brane.
The gravitational energy is abbreviated in Eq.(\ref{DBI_D4}) for simplicity.
Fifth component of gauge fields ${\cal A}_\tau$ is also neglected in
Eq.(\ref{DBI_D4}) for its large mass 
due to the complete SUSY-breaking on the D4 brane,
discussed in the previous section.
In the rescaled Yang-Mills theory with 
$g_{\rm YM}{\cal A}_\mu\rightarrow {\cal A}_\mu$, 
the coupling constant appears 
in the action as $\frac1{2g_{\rm YM}^2}\mbox{tr}{\cal F}_{\mu\nu}^2$.
Therefore, by comparing with the action (\ref{DBI_D4}),
$g_{\rm YM}$ can be written by the string parameters as
\begin{eqnarray}
g_{\rm YM}^2=\frac{g_s}{T_4 (2\pi\alpha^{'})^2 \delta\tau}
        =(2\pi)^2 g_s l_s /\delta\tau. \label{g_YM_deltatau}
\end{eqnarray}

By inverting the relations (\ref{M_KK_deltatau}) and
(\ref{g_YM_deltatau}) and also using Eq.(\ref{R_def}),
the string parameters $R$, $U_{\rm KK}$, and $g_s$ can be 
written by those in the Yang-Mills theory side as follows:
\begin{eqnarray}        
&&R^3=\frac{1}{2}\frac{g_{\rm YM}^2 N_c l_s^2}{M_{\rm KK}},\ 
U_{\rm KK}=\frac{2}{9}g_{\rm YM}^2 N_c M_{\rm KK}l_s^2,\nonumber \\
&&\hspace{18mm}g_s=\frac{1}{2\pi}\frac{g_{\rm YM}^2}{M_{\rm KK} l_s}.\label{St_Ga1}
\end{eqnarray}

Now, we see the conditions for the reliability of 
classical supergravity description.
First, it is required that the curvature everywhere is sufficiently small
relative to the fundamental string tension, so that higher-derivative
string corrections can be neglected.
Secondly, local string coupling $e^{\phi}$ should be small to suppress
the string loop effects.
These conditions give the constraints for the string parameters
and, using the relations (\ref{St_Ga1}),
they can be expressed in terms of the Yang-Mills theory,
expected to be dual with classical supergravity solution, as follows~\cite{KMMW}:
\begin{eqnarray}   
g_{\rm YM}^4 \ll \frac{1}{g_{\rm YM}^2 N_c}\ll 1, \label{sugura_reli}
\end{eqnarray}
which is achieved by $g_{\rm YM}\rightarrow 0$, $N_c\rightarrow \infty$, 
and $\lambda\equiv g_{\rm YM}^2 N_c$ fixed and large.
$\lambda$ is the 'tHooft coupling, appearing as the effective
coupling constant in the large-$N_c$ gauge theory.
Therefore, massless large $N_c$ QCD in the strong coupling regime
can be dual with classical supergravity description of D4 branes
with D8-$\overline{\rm D8}$ probes.
In other words,
nonperturbative property of QCD can be successfully described by 
the tree-level string theory.

\begin{figure}[t]
  \begin{center}
       \resizebox{70mm}{!}{\includegraphics{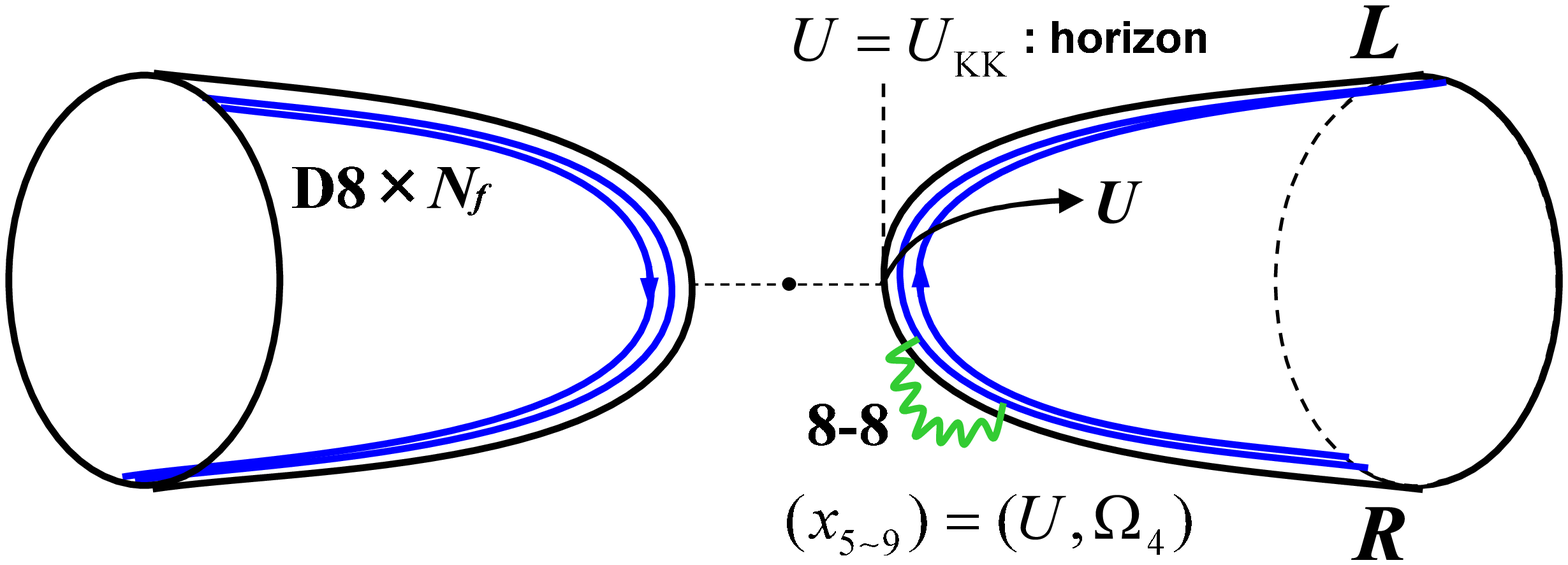}}
  \end{center}
\caption{{\small 
Probe D8 branes with D4 supergravity background (\ref{D4_sugra}).
Flat four-dimensional space-time $x_{0\sim 3}$ are not plotted.
The radial coordinate $U$ in the extra five dimensions $x_{5\sim 9}$
is bounded from below by a horizon $U_{\rm KK}$ as $U\geq U_{\rm KK}$.
Color-singlet mesons appear as the fluctuation modes of 8-8 string shown by the waving line.
The two `corns' shown above are continuously connected along the angle
directions $\Omega_4$ in $x_{5\sim 9}$.
For the argument of ${\rm SO}(5)$ singlet modes around $\Omega_4$,
only single `corn' is to be considered, as shown in Fig.3.
}}
\label{D8_backD4_1_light}
\end{figure}

Here, we comment about the realization of chiral-symmetry breaking 
and color confinement in the holographic model.
With the supergravity background of D4 brane (\ref{D4_sugra}),
probe D8 and $\overline{\rm D8}$ branes are smoothly interpolated
with each other 
in this curved space-time
as in Fig.\ref{D8_backD4_1_light} 
(see D8 brane configuration (\ref{D8_conf}) in subsection \ref{Non}),
and ${\rm U}(N_f)_{\rm D8}\times {\rm U}(N_f)_{\overline{\rm D8}}$
gauge symmetry breaks into the single ${\rm U}(N_f)$
gauge symmetry,
which is understood as a holographic manifestation of
chiral symmetry breaking in QCD.
Actually global chiral symmetry still remains as the element of 
${\rm U}(N_f)$ local gauge symmetry discussed in later sections,
and the interpolation of D8-$\overline{\rm D8}$ branes
induces its spontaneous breaking in a vacuum state.
Chiral symmetry breaking is thus realized by the geometrical connection
of D8 and $\overline{\rm D8}$ branes.

Furthermore, 
color confinement is realized in the holographic QCD
as follows.
Since color quantum number is carried only by  
$N_c$-folded D4 branes,
colored objects appear as the fluctuation modes of
open strings with at least one end located
on $N_c$-folded D4 branes, {\it e.g.}, 
gluons from 4-4 strings and flavor quarks from 4-8 strings.
Thus these colored modes are localized around the origin of 
the extra five dimensions: $x_{\alpha=5\sim 9}=0$, 
where the D4 brane exists.
Therefore, in the supergravity background of D4 brane,
it can be interpreted that these colored objects
would locate behind the horizon $U_{\rm KK}$
and become invisible from outside,
which can be understood as the color-confinement phenomena at the low-energy scale.
In fact, in the holographic QCD, color confinement is 
realized as `loss of direct color information' at the low-energy scale, 
due to the appearance of `non-trivial large spacial curvature'  
in the extra fifth-direction as shown in Fig.\ref{D8_backD4_1_light}, 
while the informations from color degrees of freedom are hidden at large distances 
due to extremely strong correlations in usual four-dimensional QCD.
The appearance of the horizon in the classical supergravity solution, 
which is reliable for large $N_c$ 
and large 't~Hooft coupling $\lambda$, 
seems consistent with the low-energy property of large-$N_c$ Yang-Mills
theory, inducing self-coupling of gauge fields and the confinement. 

These considerations may suggest that
chiral symmetry breaking and color confinement occur simultaneously;
two independent chirality spaces are connected by the `worm hole' into
which colored objects are absorbed as shown in Fig.\ref{D8_backD4_1_light}.

The close relations between chiral symmetry breaking and color confinement
have been observed in lattice QCD calculations at finite temperatures~\cite{R05}.
In lattice QCD at finite temperatures,
chiral symmetry breaking is expressed 
by the quark condensate $\langle \bar qq \rangle={\rm tr} G_q[{\cal A}_\mu]$
as its order parameter
with the light quark propagator $G_q[{\cal A}_\mu]=\frac{1}{\not p+g \not
{\cal A}+m_q}$ in the Euclidean metric.
The quark confinement property can also be expressed 
by the vacuum expectation value of the Polyakov-loop 
$\langle P \rangle$ with $P \equiv {\rm tr} {\rm P} e^{ig\int_0^{1/T}
d\tau {\cal A}_4({\bf x},\tau)}$, 
which is the order parameter of the center symmetry ${\bf Z}^{N_c} \in {\rm SU(N_c)}$  
and is physically related to the single static-quark energy $E_q$ as $\langle P \rangle=e^{-E_q/T}$.
In fact, the chiral-symmetry restoration and the deconfinement phase transition 
are found to occur
simultaneously at the critical temperature $T_c$ in both quenched and full lattice QCD with 
$N_c$=2,3 and $N_f$=0,1,2,3,4: both of the order parameters, {\it i.e.}, the Polyakov loop $\langle P \rangle$ and 
the quark condensate $\langle \bar qq \rangle$, drastically change around $T_c$ near the chiral limit.

In spite of the lattice QCD evidence, the relations are still unclear 
between chiral symmetry breaking and color confinement,
especially from the analytical frameworks. 
Furthermore, at finite baryon-number density, which is also an important axis in the QCD phase diagram,  
lattice QCD calculations do not work well due to the `sign problem'.
Therefore 
there is still missing the sufficient consensus about the
relations of chiral symmetry breaking and color confinement, 
particularly in the finite-density QCD.
From these viewpoints, it would be desired to perform the well-defined generalization of 
the holographic models to the finite-temperature and finite-density QCD in future.

After the supergravity description of background D4 branes,
there only appear colorless objects as the fluctuation modes of
residual probe D8 branes:
mesons and also baryons.
By the analysis of large-$N_c$ gauge theory~\cite{tHooft},
large-$N_c$ QCD is found to be equivalent with the 
weak-coupling system of mesons (and glueballs), and 
baryons do not directly appear as the dynamical degrees of freedom.
Therefore, in this large-$N_c$ holographic QCD,
baryons are expected to appear not directly but
as some soliton-like topological objects,
which is discussed in later sections.
 
\subsection{Non-abelian DBI action of probe D8 brane \label{Non}}
%
In this section, we treat the multi-flavor system with $N_f=2$,
by using 
the non-abelian Dirac-Born-Infeld (DBI) action of D8 brane as
\begin{eqnarray}
\hspace*{-2mm}S_{\rm D8}^{\rm DBI} \!\!&=&\!\!
 T_8\int d^9 x e^{-\phi}\sqrt{-\det\lk g_{MN}+2\pi \alpha^{'}
 F_{MN}\rk}\label{DBI_1}\\
 &=&\!\!
  T_8\int d^9 x e^{-\phi}\sqrt{-\det g_{MN}}\nonumber \\%
  &&\!\!\times \ltk 1+\frac{1}{4}(2\pi \alpha^{'})^2%
               g^{JK}g^{PQ} 2\mbox{tr}\lk
	       F_{JP}F_{KQ}\rk\rtk \nonumber \\
  &&\!\!+O(F^4),\label{DBI_2}
\end{eqnarray}
where $F_{MN}=\partial_M A_N-\partial_N A_M+i\ldk A_M,A_N\rdk $ is the field
strength tensor in nine-dimensional space-time of D8 brane,
and $A_{\mu=0\sim 3}$, $A_z$, and $A_{\alpha=5\sim 8}$
($\alpha=5\sim 8$ are coordinate indices on $S^4$) are 
the hermite ${\rm U}(N_f)$ gauge fields 
as the fluctuation modes of 8-8 string.
Factor $2$ appears in the second term of (\ref{DBI_2}) according to the
normalization of generators: $\mbox{tr}T^a T^b=\frac{1}{2}\delta^{ab}$. 
In fact, gauge field $A_M$ is color-singlet and 
obeys the adjoint representation of ${\rm U}(N_f)$ flavor space,
eventually producing the meson degrees of freedom in the holographic QCD. 

\begin{figure}[t]
  \begin{center}
       \resizebox{87mm}{!}{\includegraphics{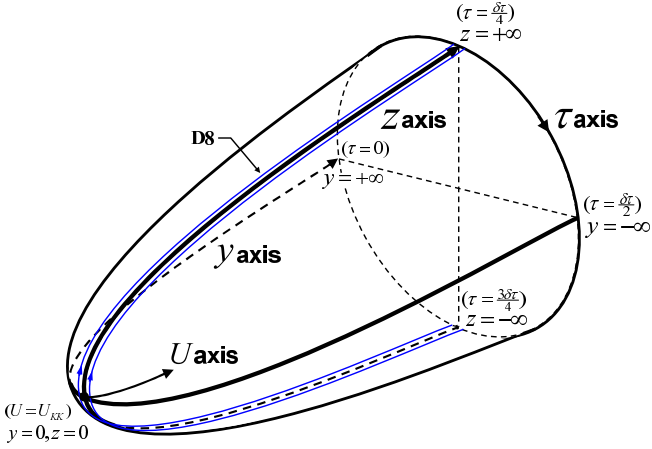}}\\
  \end{center}
\caption{ {\small Coordinates $(U, \tau)$ and $(y, z)$ on the D4 supergravity
solution (\ref{D4_sugra});
coordinates $(y, z)$ are introduced as 
$\frac{y}{U_{\rm KK}}\equiv\lk\frac{U^3}{U_{\rm
 KK}^3}-1\rk^{\frac{1}{2}}\cos\lk \frac{2\pi}{\delta\tau}\tau\rk$  and
$\frac{z}{U_{\rm KK}}\equiv\lk\frac{U^3}{U_{\rm
 KK}^3}-1\rk^{\frac{1}{2}}\sin\lk \frac{2\pi}{\delta\tau}\tau\rk$.
Stabilized D8 brane configuration (\ref{D8_conf}) with the curved
 space-time background is also shown.} }
\label{D8_coodinates_light}
\end{figure}

In $A_M\rightarrow 0$ limit, the action (\ref{DBI_2}) becomes
\begin{eqnarray}
S_{\rm D8}^{\rm DBI}|_{A_M\rightarrow 0}=T_8\int d^9 x e^{-\phi}\sqrt{-\det
 g_{MN}},\label{DBI1_limit}
\end{eqnarray}
which gives the gravitational energy of D8 brane in general
coordinates.
By minimizing the action (\ref{DBI1_limit}) with D4 supergravity
background (\ref{D4_sugra}),
the configuration of probe D8 brane in ten-dimensional space-time 
is determined~\cite{SS} to be
\begin{eqnarray}
\tau(U)=\delta\tau/4,\label{D8_conf}
\end{eqnarray}
which is shown in Fig.\ref{D8_backD4_1_light},
and coordinates on the D4 supergravity solution in detail
with the stabilized D8 brane configuration (\ref{D8_conf})
are shown in Fig.\ref{D8_coodinates_light}.

Now, we discuss the gauge fields as the fluctuation modes of D8 brane
configuration (\ref{D8_conf}).
${\rm SO}(5)$ rotational symmetry around $S^4$ does not exist in QCD
with four-dimensional space-time.
Therefore, in the D4/D8/$\overline{\rm D8}$ holographic model, 
only ${\rm SO}(5)$ singlet modes are considered
and gauge fields $A_{\alpha=5\sim8}$ are neglected as ${\rm SO}(5)$
vectors.
$A_\mu$ and $A_z$ are also assumed to be independent of these extra
coordinates on $S^4$.
Then the integral along the $S^4$ direction in the action (\ref{DBI_2}) can be performed as
\begin{eqnarray}
 &&\hspace{-5mm}S_{\rm D8}^{\rm DBI}-S_{\rm D8}^{\rm DBI}|_{A_M\rightarrow 0} \nonumber \\
 &&\hspace{-5mm}= \tilde{T}_8(2\pi\alpha^{'})^2\int d^4x dz\ 2\mbox{tr} \bigg\{
   \frac{R^3}{4U_z(z)}g^{\mu\nu}g^{\rho\sigma}F_{\mu\rho}F_{\nu\sigma}
   \nonumber \\
   &&\hspace{-1mm}
    +\frac{9}{8}\frac{U_z^3(z)}{U_{\rm KK}}g^{\mu\nu}F_{\mu z}F_{\nu
   z}\bigg\} + O(F^4),\label{5dimDBI}
\end{eqnarray}
where 
$\tilde{T}_8\equiv \frac{2}{3}R^{\frac{3}{2}}U_{\rm KK}^{\frac{1}{2}}T_8V_4 g_s^{-1}$
and the fifth coordinate $z$ on the D8 brane configuration (\ref{D8_conf})
and the $z$-dependent function $U_z(z)$ are defined as
\begin{eqnarray}
\frac{z^2}{U_{\rm KK}^2} \!&\equiv&\! \frac{U^3}{U_{\rm KK}^3}-1,
\label{z_def}\\
U_z(z) \!&\equiv&\! \lk U_{\rm KK}^3 +U_{\rm KK}z^2
 \rk^{\frac{1}{3}}.\label{Uz_def}
\end{eqnarray}
In the action (\ref{5dimDBI}), the gravitational energy (\ref{DBI1_limit}) is subtracted 
as a vacuum energy relative to the gauge sectors.
The curvature in the fifth dimension is shown
by the induced measures $\frac{R^3}{4U_z}$ and
$\frac{9}{8}\frac{U_z^3}{U_{\rm KK}}$ in Eq.(\ref{5dimDBI});
$F_{\mu z}$ includes a field strength along the $z$ direction, 
so that the measure for $F_{\mu z}$ differs from that for $F_{\mu\nu}$.
The five-dimensional Yang-Mills theory (\ref{5dimDBI})
is regarded as the `unified' theory of mesons,
where mode expansion of the gauge fields 
gives the meson degrees of freedom, discussed in later sections.

Now one can easily show that, using the relations (\ref{St_Ga1}),
DBI action of D8 brane (\ref{5dimDBI}) with D4 supergravity
background 
has no explicit dependence on the string length $l_s$.
This is because the effective action with classical supergravity
background of D4 brane is constructed with the condition 
$U_{\rm KK}^{\frac{1}{2}}R^{\frac{3}{2}}\gg l_s^2$
($U_{\rm KK}^{\frac{1}{2}}R^{\frac{3}{2}}$ is the typical curvature
in D4 classical supergravity solution~\cite{KMMW}), {\it i.e.},
the reliability of the local approximation for the
fundamental strings,
which gives the large 'tHooft coupling condition
$\lambda \gg 1$ in the gauge theory side discussed in
Eq.(\ref{sugura_reli}).
Therefore, without loss of generality,
we can set the string tension as sufficiently small value 
$l_s^2=\frac{9}{2}\lambda^{-1}M_{\rm KK}^{-2}$
to be consistent with the reliability of the local approximation for the 
fundamental strings.
Then the relations between the dimensional quantities
$R$, $U_{\rm KK}$, and
$M_{\rm KK}$ in Eqs.(\ref{St_Ga1})
can be rewritten as
\begin{eqnarray}
R^3=\frac{9}{4}M_{\rm KK}^{-3},\hspace{5mm}
U_{\rm KK}=M_{\rm KK}^{-1}.\label{unit_set_UM}
\end{eqnarray}
In this set up, all the dimensional quantities are simply scaled 
with the Kaluza-Klein mass $M_{\rm KK}$, and therefore 
we take the convenient unit $M_{\rm KK}=1$ for the derivation of the 
four-dimensional effective theory.
At the end of calculations, the dimensional quantities can be recovered 
by multiplying the proper power of $M_{\rm KK}$.

The final form of the DBI action in $M_{\rm KK}=1$ unit is written 
as the five-dimensional Yang-Mills theory 
with flat four-dimensional Euclidean space-time $x$, {\it i.e.}, $x_{0\sim 3}$
and other fifth dimension $z$ with curved measure as
\begin{eqnarray}
&&\hspace{-2mm}S_{\rm D8}^{\rm DBI}-S_{\rm D8}^{\rm DBI}|_{A_M\rightarrow 0} \nonumber \\
&&\hspace{-2mm} =\kappa\int d^4 x dz\mbox{tr}\ltk
\frac{1}{2}K(z)^{-\frac{1}{3}}F_{\mu\nu}F_{\mu\nu}+K(z) F_{\mu z}F_{\mu
z}\rtk \nonumber \\
&&\hspace{1.4mm} +O(F^4),
\label{5dimDBI_2}
\end{eqnarray}
where the overall factor $\kappa$ is defined in the $M_{\rm KK}=1$ unit as
\begin{eqnarray}
\kappa\equiv \tilde{T}_8(2\pi\alpha^{'})^2 R^3=\frac{\lambda N_c}{108
\pi^3},\label{kappa_def}
\end{eqnarray}
and
$K(z)\equiv 1+z^2$ expresses the nontrivial curvature 
in the fifth direction $z$ induced by the supergravity 
background of the D4 brane.

In the $M_{\rm KK}=1$ unit,
the factor $\alpha^{'}(=l_s^2)$ appearing in front of the field strength
tensor in the DBI action (\ref{DBI_1})
can be written as $\alpha^{'}=\frac{9}{2}\lambda^{-1}$.
Thus it becomes clear that 
the expansion of the DBI action with respect to the field strength
tensor corresponds to the expansion about the `inverse' of the large 'tHooft coupling
$\lambda$ with condition (\ref{sugura_reli}),
where
the non-perturbative aspects of QCD with large 'tHooft coupling $\lambda$
can be manifestly linked with the tree-level Yang-Mills theory
induced on the probe D8 brane with D4 supergravity background.
This is similar to the non-linear sigma model~\cite{Wein} 
as the meson effective theory 
having soft-momentum expansion,
which gives a workable perturbative treatment in the strong-coupling regime
of QCD.
We now treat 
non-trivial leading order $O(F^2)$ of the DBI action (\ref{5dimDBI_2}),
corresponding to the leading of $1/N_c$ and $1/\lambda$ expansions
in the holographic QCD,
for the argument of the non-perturbative (strong coupling) properties of
QCD.

\subsection{Mode expansion of gauge field \label{Mod}}
%
In this section, to obtain the four-dimensional effective theory
reflecting definite parity and G-parity of QCD,
we perform the mode expansion of the gauge field with respect to
the four-dimensional parity eigen-states 
by using proper orthogonal basis in the $z$ direction.

The gauge field $A_M(x_N)$ ($M,N=0\sim 3,z$) transforms with ${\rm U}(N_f)$
local gauge symmetry of D8 brane as
\begin{eqnarray}
\hspace{-5mm}A_M(x_N)\rightarrow A_M^g(x_N)\!\! 
&\equiv&\!\! g(x_N) A_M(x_N) g^{-1}(x_N)
 \nonumber \\
 &&\!\!+\frac{1}{i}g(x_N)\partial_M g^{-1}(x_N).\label{gauge_tra1}
\end{eqnarray}

In order that the action (\ref{5dimDBI_2}) is to be finite
as the tree level,
the field strength tensor $F_{MN}$ should vanish at 
$z\rightarrow \pm\infty$.
This can be achieved by the gauge field $A_{\mu}$
which goes to the pure-gauge form in $z\rightarrow \pm\infty$
as $A_M\rightarrow (1/i)U^{\dagger}\partial_M U$ ($U\in U(N_f)$).
Now, by applying the gauge-transformation (\ref{gauge_tra1}), 
we begin with the gauge fixing satisfying
\begin{eqnarray}
A_M(x_N)\rightarrow 0. \hspace{6mm} 
\mbox{(as $z\rightarrow \pm\infty$)}\label{bound1}
\end{eqnarray}
In this gauge fixing, the gauge degrees of freedom
is fixed only at the 4-dimensional `surfaces' 
$z=\pm\infty$ in five-dimensional space-time, 
and there still remains large gauge-degrees of freedom with
gauge function $g^{res}(x_M)$ as
\begin{eqnarray}
\partial_M g^{res}(x_N) \rightarrow 0, \hspace{5mm} 
\mbox{(as $z\rightarrow \pm\infty$)}\label{chiral1}
\end{eqnarray}
which means that $g^{res}(x_M)$ goes to constant matrix $g_{\pm}\in U(N_f)$ as 
$g_\pm\equiv {\displaystyle \lim_{z\rightarrow\pm\infty}}g^{res}(x_M)$.
In the holographic model, ($g_+,g_-$) are regarded as the 
elements of chiral symmetry ${\rm U}(N_f)_L\times {\rm U}(N_f)_R$ in QCD;
the global chiral symmetry $(U(N_f)_L, U(N_f)_R)$ appears as the elements 
of the local $U(N_f)$ gauge symmetry of D8 brane at the surfaces
$z=\pm \infty$ in the 
five-dimensional space-time.
Thus, in the five-dimensional description, 
$U(N_f)_L$ and $U(N_f)_R$ appear and act at spatially different surfaces
$z=\pm\infty$,
by holding local ${\rm U}(N_f)$ gauge symmetry between two of them.

Now, pion field is introduced in refs.~\cite{SS, DTSon} as 
\begin{eqnarray}
U(x^{\mu})=P \exp\ltk
 -i\int_{-\infty}^{\infty}dz^{'}A_z(x_\mu,z^{'})\rtk \in U(N_f),
\label{pion1}
\end{eqnarray}
where $P$ denotes the path-ordered product along the $z$ coordinate
with the upper-edge $z$ variable arranged in left-hand side.
Since $g^{res}(x_M)$ is $x$-independent at $z=\pm \infty$, 
$U(x^\mu)$ transforms by the gauge function $g^{res}(x_M)$ 
as the global transformation as
\begin{eqnarray}
U(x^{\mu}) \!\!&\rightarrow& \!\!g^{res}(x_\mu, z=+\infty) U(x^{\mu}) 
                      g^{res}(x_\mu, z=-\infty)^{-1} \nonumber\\
 &=&\!\!  g_+ U(x^{\mu}) g_-^{-1}.\label{chiral_tra}
\end{eqnarray}
This transformation property is the same as that for the pion field in the chiral Lagrangian of
non-linear sigma model with respect to global chiral
transformation~\cite{SS, DTSon, Wein}.

Here, we also introduce the variables $\xi_\pm(x_\mu)$ as
\begin{eqnarray}
\xi_{\pm}^{-1}(x^{\mu})\!=\!P\exp\ltk \!-i\int_{z_0(x_\mu)}^{\pm\infty}
\!\!\!\! dz^{'}A_z(x_\mu,z^{'})\rtk\! \in U(N_f),\label{xi1}
\end{eqnarray}
where $z_0(x_\mu)$ is a 
single-valued arbitrary function of $x_\mu$.
Then the chiral field (\ref{pion1}) can be written as
\begin{eqnarray}
U(x^{\mu})=\xi_{+}^{-1}(x_\mu)\xi_-(x_\mu).\label{pion2}
\end{eqnarray}
$\xi_\pm(x_\mu)$ transform with the gauge function $g^{res}(x_M)$ as
\begin{eqnarray}
\xi_{\pm}(x_\mu)\rightarrow h_{z_0}(x_\mu)\xi_{\pm}(x_\mu)g_{\pm}^{-1},\label{HLS_z0}
\end{eqnarray}
where $h_{z_0}(x_\mu)\equiv g^{res}(x_\mu, z=z_0(x_\mu))$ is the element of
${\rm U}(N_f)$ local gauge symmetry on the four-dimensional surface $z_0(x_\mu)$.
This local gauge symmetry of $h_{z_0}(x_\mu)$ corresponds to the 
hidden local symmetry for vector mesons 
in ref.~\cite{BKY}, which is later discussed.

Now, we change to the $A_z=0$ gauge,
which is a popular gauge in the holographic models.
The $A_z=0$ gauge is appropriate for the construction of 
the four-dimensional effective theory because
$A_z$ behaves as a scalar field in the flat four-dimensional space-time
and it is to be absorbed
by the four-dimensional gauge field $A_\mu$ (see section 3.4 of
the former reference in ~\cite{SS}).
In fact, in the $A_z=0$ gauge,  
the breaking of ${\rm U}(N_f)$ local gauge symmetry
becomes manifest
like the unitary gauge in the non-abelian Higgs theory,
which eventually induces the finite mass for gauge fields,
especially for the (axial) vector mesons appearing as a part of $A_\mu$. 
One can take the $A_z=0$ gauge by applying gauge function 
\begin{eqnarray}
g^{-1}(x_{\mu}, z)=P\exp\ltk
 -i\int_{z_0(x_\mu)}^{z}dz^{'}A_z(x_\mu,z^{'})\rtk,\label{Az0gauge}
\end{eqnarray}
which changes the boundary behavior of $A_\mu$ as
\begin{eqnarray}
A_\mu(x_N)\rightarrow
 \frac{1}{i}\xi_{\pm}(x_\nu)\partial_\mu\xi_{\pm}^{-1}(x_\nu).\hspace{5mm} 
\mbox{(as $z\rightarrow \pm\infty$)}\label{limit1}
\end{eqnarray}

Now, we perform the mode expansion of $A_\mu$ with the boundary condition
(\ref{limit1}) by using proper orthogonal basis
$\psi_\pm(z)$ and $\psi_{n}(z)$ $(n=1,2,\cdots)$ as follows:
\begin{eqnarray}
&&\hspace{-9.5mm}A_\mu(x_N) \nonumber \\
&&\hspace{-9.5mm}=l_\mu(x_\nu)\psi_+(z)+
   r_\mu(x_\nu)\psi_-(z)\!+\!
 \sum_{n\geq 1}B_\mu^{(n)}(x_\nu)\psi_n(z),\label{limit2_lr}\\
&&\hspace{8mm} l_\mu(x_\nu)\equiv
 \frac{1}{i}\xi_{+}(x_\nu)\partial_\mu\xi_{+}^{-1}(x_\nu),\\ 
&&\hspace{8mm}
 r_\mu(x_\nu)\equiv
 \frac{1}{i}\xi_{-}(x_\nu)\partial_\mu\xi_{-}^{-1}(x_\nu).
 \label{lr_def}
\end{eqnarray}
%
The basis $\psi_\pm$ are chosen to support all parts of the gauge field $A_\mu$ at
$z\rightarrow \pm\infty$ as 
$\psi_\pm(z\rightarrow \pm\infty)=1$ and
$\psi_\pm(z\rightarrow \mp\infty)=0$,
which are here defined as 
\begin{eqnarray}
\psi_\pm(z)\!\!&\equiv&\!\! 
\frac{1}{2}\pm\hat{\psi}_0(z),\label{hat_psi_0}\\
\hat{\psi}_0(z)\!\!&\equiv&\!\!
\frac{1}{\pi}\arctan z.\label{psi_n}
\end{eqnarray}
In order to diagonalize the DBI action (\ref{5dimDBI_2}) with 
the curved measures $K(z)^{-\frac{1}{3}}$ and $K(z)$ induced in the fifth direction, 
the basis $\psi_n$ $(n=1,2,\cdots)$ are taken to be the normalizable eigen-function satisfying
\begin{eqnarray}
-K(z)^{\frac{1}{3}}\frac{d}{dz}\ltk K(z) \frac{d\psi_n}{dz}\rtk =
 \lambda_n\psi_n,
\hspace{2mm}\mbox{($\lambda_1 < \lambda_2 < \cdots$)}\label{eigen1}
\end{eqnarray}
with normalization condition as 
\begin{eqnarray}
\kappa\int dz K(z)^{-\frac{1}{3}}\psi_m \psi_n =\delta_{nm}.\label{normal_psi_n}
\end{eqnarray}
It should be noted that $\psi_\pm$, {\it i.e.}, $\hat{\psi}_0$ 
in Eqs.(\ref{hat_psi_0}) and (\ref{psi_n}) can be regarded 
as the zero-modes of (\ref{eigen1}) with eigen-value $\lambda_0\equiv 0$
as 
\begin{eqnarray}
-K(z)^{\frac{1}{3}}\frac{d}{dz}\ltk K(z) \frac{d\psi_{\pm}}{dz}\rtk =0,\label{zero_mode_con1}
\end{eqnarray}
although they are not normalizable.

Eqs.(\ref{hat_psi_0})$\sim$(\ref{normal_psi_n}) also give other
conditions for the basis with $z$-derivatives as
\begin{eqnarray}
\kappa\int dz K(z)\partial_z\psi_n\partial_z\psi_m
    \!\!&=&\!\!\lambda_n\delta_{nm},\label{ortho_deri1}\\
\kappa\int dz K(z)\partial_z\hat{\psi}_0\partial_z\psi_m
    \!\!&=&\!\!\frac{\kappa}{\pi}\int dz \partial_z\psi_m=0.\label{ortho_deri0}
\end{eqnarray}
Conditions (\ref{normal_psi_n}$\sim$\ref{ortho_deri0}) for orthogonal basis 
are used to give the well-defined
mode decoupling of the action, which is later discussed in section \ref{Eff}.

In the holographic model, the fields $B_\mu^{(n=1,2,\cdots)}$ in
Eq.(\ref{limit2_lr}) are identified as the (axial) vector mesons,
which belong to the adjoint representation of the $U(N_f)$ 
gauge group as
$B_\mu=B_\mu^aT_a$.
After diagonalizing the DBI action with infinite number of (axial)
vector mesons $B_\mu^{(n=1,2,\cdots)}$, 
their mass terms appear with $m_n^2\equiv \lambda_n$~\cite{SS}.
Note here that, in the mode expansion of the gauge field (\ref{limit2_lr}), 
$A_M(x_N)$ is a five-dimensional vector field obeying 
$A_M(x_N) \rightarrow -A_M(x_N)$ under the 
five-dimensional parity transformation,  
and we can choose $\psi_n$ as the parity eigen-states of (\ref{eigen1}) as
$\psi_n(-z)=(-)^{n-1}\psi_n(z) $ 
using the translational invariance along $z$-direction.
Then, $B_\mu^{(n=1,2,\cdots)}$ transform for the 
four-dimensional 
parity transformation as $B_\mu^{(n)}(x_\nu) \rightarrow (-)^nB_\mu^{(n)}(x_\nu)$.
These considerations show that
vector and axial vector mesons appear alternately in the 
excitation spectra~\cite{SS}.

Even with $A_z=0$ gauge fixing, there still remains residual gauge
symmetry appearing in (\ref{HLS_z0}) as  
\begin{eqnarray}
A_\mu(x_N)\!\!&\rightarrow&\!\! h_{z_0}(x_\nu)A_\mu(x_N) h_{z_0}^{-1}(x_\nu)
 \nonumber \\
                      &&+\frac{1}{i}h_{z_0}(x_\nu)\partial_\mu h_{z_0}^{-1}(x_\nu),\label{hidden_A}\\
B_\mu^{(n)}(x_\nu)\!\!&\rightarrow&\!\! h_{z_0}(x_\nu)B_\mu^{(n)}(x_\nu)
   h_{z_0}^{-1}(x_\nu).\label{hidden_B}
\end{eqnarray}
%
Note that there also exist the symmetry 
of deforming the surface $z_0(x_\mu)$ to the $z$-direction, 
apart from the gauge symmetry as in Eqs.(\ref{hidden_A}) and (\ref{hidden_B}).
This means that there exist infinite number of
the hidden local symmetries for infinite number of (axial) vector mesons
with respect to the surface degrees of freedom.
This is similar to the `open moose model' discussed by Son {\it et al.}~\cite{DTSon},
which is the phenomenological five-dimensional meson theory constructed
by embedding infinite number of hidden local symmetries in the chiral field 
defined on the flat four-dimensional space-time.

Eq.(\ref{limit2_lr}) can also be written by using useful variables $\alpha_\mu$ and
$\beta_\mu$ as
\begin{eqnarray}
&&\hspace{-8mm}A_\mu(x_N) \nonumber \\
&&\hspace{-8mm}=\alpha_\mu(x_\nu)\hat{\psi}_0(z) +
           \beta_\mu(x_\nu)+
           \sum_{n\geq 1}B_\mu^{(n)}(x_\nu)\psi_n(z),
\end{eqnarray}
%
%
where $\alpha_\mu$ and $\beta_\mu$ are defined as
\begin{eqnarray}
&&\alpha_\mu(x_\nu)\equiv l_\mu(x_\nu)-r_\mu(x_\nu), \label{AB_add} \\
&&\beta_\mu(x_\nu) \equiv \frac{1}{2}\ltk l_\mu(x_\nu)+r_\mu(x_\nu)\rtk.
\label{AB_def}
\end{eqnarray}
$\alpha_\mu$ and $\beta_\mu$ correspond to 
`axial vector' and `vector' current, respectively.
With the residual gauge symmetry
with the gauge function $h_{z_0}(x_\mu)$ in Eqs.(\ref{hidden_A}) and (\ref{hidden_B}),
$\alpha_\mu$ and $\beta_\mu$ transform as
\begin{eqnarray}
\alpha_\mu\rightarrow h_{z_0}\alpha_\mu h_{z_0}^{-1},\hspace{3mm}
 \beta_\mu\rightarrow h_{z_0}\beta_\mu h_{z_0}^{-1}+\frac{1}{i}h_{z_0}\partial_\mu
 h_{z_0}^{-1}.\label{ab_tra}
\end{eqnarray}

Now we take $\xi_{+}^{-1}(x_\mu)=\xi_{-}(x_\mu)$ gauge for
the residual gauge symmetry in Eqs.(\ref{hidden_A}) and
(\ref{hidden_B}).
%
%
This type of gauge fixing is used in the hidden local symmetry 
approach~\cite{BKY}, which gives the breaking of hidden local
symmetry ${\rm SU} (N_f)_V$ and generates the mass of vector mesons as 
the Higgs mechanism. 
The chiral field $U(x_\mu)$ is written in analogy with that of the chiral Lagrangian as
$U(x_\mu)=e^{2i\pi(x_\mu)/f_\pi}\in {\rm U}(N_f)$ with pion field
$\pi(x_\mu)$.
Then, in the $\xi_{+}^{-1}(x_\mu)=\xi_{-}(x_\mu)$ gauge,
$\xi_{\pm}$ can be written from Eq.(28) as
\begin{eqnarray}
\xi_{+}^{-1}(x_\mu)=\xi_{-}(x_\mu)\equiv\xi(x_\mu)=e^{i\pi(x_\mu)/f_\pi}.\label{pm_gauge}
\end{eqnarray}
With this gauge fixing (\ref{pm_gauge}), 
$\xi_{\pm}$ become
parity and G-parity partners with each other
in four-dimensional space-time,
and $\alpha_\mu$ and $\beta_\mu$ transform as axial vector and vector fields.                
Therefore, if the action in the $\xi_{+}^{-1}(x_\mu)=\xi_{-}(x_\mu)$
gauge is parity or G-parity transformed
in four-dimensional space-time,
then gauge-fixing condition shifts into 
$\xi_{-}^{-1}(x_\mu)=\xi_{+}(x_\mu)$,
which is the same gauge fixing as the previous one.
In this sense, the effective action in the $\xi_{+}^{-1}(x_\mu)=\xi_{-}(x_\mu)$
gauge is realized as a scalar ({\it i.e.,} invariant) for the four-dimensional parity and
G-parity transformation.
In fact, other gauge, {\it e.g.}, $\xi_-(x_\mu)=1$ gauge gives $\rho$-$3\pi$
couplings like 
$(\partial_\mu \mbox{\boldmath $\pi$})^2 (\partial_\nu \mbox{\boldmath
$\pi$}\cdot \mbox{\boldmath $\rho$}_\nu)$
and $(\partial_\mu\mbox{\boldmath $\pi$}\cdot\mbox{\boldmath
$\rho$}_\nu)(\partial_\mu\mbox{\boldmath
$\pi$}\cdot\partial_\nu\mbox{\boldmath $\pi$})$
in the four-dimensional effective action,
which violate both parity and G-parity.
In terms of the parity and G-parity classification in the strong interaction, 
it is essential to take the $\xi_+^{-1}(x_\mu)=\xi_-(x_\mu)$ gauge 
to construct four-dimensional effective theory of QCD.

\subsection{Effective action of pion and $\rho$ meson \label{Eff}}
%
In this paper, we treat pions together with $\rho$ mesons 
as the lightest vector mesons.
We neglect higher mass excitation modes of the vector (axial vector)
mesons in discussing the low-energy properties of QCD.
Especially as for the profiles of chiral solitons,
the effects of heavier (axial) vector mesons are expected to be small
with the small coupling constants between
pions and heavier (axial) vector mesons,
which is discussed in the end of this subsection.
Now, the $\rho$ meson field is introduced as 
$\rho_\mu\equiv B_\mu^{(1)}$, which couples with the lowest mode
$\psi_1(z)$ in the fifth direction.
Then the gauge field can be written as
\begin{eqnarray}
&&\hspace{-7.3mm}A_\mu(x_{\nu},z) \nonumber \\
&&\hspace{-7.3mm}=l_\mu(x_\nu)\psi_+(z)+
   r_\mu(x_\nu)\psi_-(z)+
 \rho_\mu(x_\nu)\psi_1(z),\label{exp_rhomode}
\end{eqnarray}
and the five-dimensional field strength tensor can be also written
with the $A_z=0$ gauge as
%
\begin{eqnarray}
F_{\mu\nu}\!\!&=&\!\!\partial_\mu A_\nu-\partial_\nu A_\mu+
             i\ldk A_\mu,A_\nu\rdk\nonumber\\ 
          &=&\!\!\lk\partial_\mu l_\nu-\partial_\nu l_\mu \rk \psi_+ 
           +\lk\partial_\mu r_\nu-\partial_\nu r_\mu \rk \psi_-
	   \nonumber\\
          &&\hspace{-4mm} +\lk\partial_\mu \rho_\nu-\partial_\nu \rho_\mu \rk
           \psi_1\nonumber\\
         &&\hspace{-4mm}+i\ltk [l_\mu,l_\nu ]\psi_+^2+
                [r_\mu,r_\nu ]\psi_-^2+
                [\rho_\mu,\rho_\nu ]\psi_1^2
           \rtk\nonumber\\
         &&\hspace{-4mm}+i\{ \lk [l_\mu,r_\nu ]+[r_\mu,l_\nu ] \rk\psi_+\psi_- \nonumber\\  
                 &&\hspace{-1.5mm} +\lk [l_\mu,\rho_\nu ]+[\rho_\mu,l_\nu ] \rk\psi_+\psi_1 \nonumber\\   
                 &&\hspace{-1.5mm} +\lk [r_\mu,\rho_\nu ]+[\rho_\mu,r_\nu ]
		  \rk\psi_-\psi_1 
             \}\nonumber\\
         &=&\!\!-i[\alpha_\mu,\alpha_\nu]\psi_+\psi_- +
             \lk \partial_\mu\rho_\nu -\partial_\nu \rho_\mu \rk \psi_1
             +i[\rho_\mu,\rho_\nu]\psi_1^2\nonumber\\
         &&\hspace{-4mm} +i\{ 
               \lk[\alpha_\mu,\rho_\nu]+[\rho_\mu,\alpha_\nu]\rk\hat{\psi}_0\psi_1\nonumber\\ 
	       &&\hspace{-1.5mm}+\lk[\beta_\mu,\rho_\nu]+[\rho_\mu,\beta_\nu]\rk\psi_1
              \},\label{fieldst1}\\
F_{z\mu}\!\! &=&\!\!\partial_z A_\mu 
              =\alpha_\mu\partial_z\hat{\psi}_0+\rho_\mu\partial_z\psi_1.\label{fieldst2}               
\end{eqnarray}
In the derivation of Eqs.(\ref{fieldst1})$\cdot$(\ref{fieldst2}), 
we have used the Maurer-Cartan equations, 
$\partial_\mu l_\nu-\partial_\nu l_\mu+i[l_\mu,l_\nu]=0$ and 
$\partial_\mu r_\nu-\partial_\nu r_\mu+i[r_\mu,r_\nu]=0$, and
definitions of $\alpha_\mu$ and $\beta_\mu$
in Eqs.(\ref{AB_add}) and (\ref{AB_def}).

The second term of the DBI action (\ref{5dimDBI_2}) can be diagonalized by using
Eqs.(\ref{ortho_deri1})$\cdot$(\ref{ortho_deri0}) as
\begin{eqnarray}
&&\hspace{-5mm}
\kappa\int d^4x dz \mbox{tr} \ltk K(z) F_{\mu z}F_{\mu z}\rtk \nonumber \\
&&\hspace{-5mm}= \kappa\int d^4 x dz K(z){\rm tr}\lk
\alpha_\mu\partial_z\hat{\psi}_0+\rho_\mu\partial_z\psi_1\rk^2\nonumber\\ 
&&\hspace{-5mm}=\int d^4 x\ltk \frac{f_\pi^2}{4}\mbox{tr} (L_\mu L_\mu)+
                m_\rho^2 \mbox{tr} (\rho_\mu \rho_\mu) 
           \rtk,\label{DBI_2nd}\\
&&\hspace{13.7mm}
L_\mu \equiv \frac{1}{i}U^{\dagger}\partial_\mu U, \label{Lone_form}
\end{eqnarray}
where 
the relation $\alpha_\mu=\xi_-L_\mu \xi_-^{-1}$ is used.
In Eq.(\ref{DBI_2nd}),
the pion decay constant $f_\pi$
is introduced, comparing to the chiral Lagrangian, as
\begin{eqnarray}
\frac{f_{\pi}^2}{4} \!&\equiv&\!\kappa\int dz 
                           K(z) (\partial_z\hat{\psi}_0)^2
=\frac{\kappa}{\pi},\label{f_pi_1}
\end{eqnarray}
and the mass of $\rho$ meson field $\rho_\mu\equiv B_\mu^{(1)}$
is presented from the orthogonal condition (\ref{ortho_deri1}) as
\begin{eqnarray}
m_\rho^2\equiv m_1^2=\lambda_1.\label{rho_mass}
\end{eqnarray}
%
%
%

The first term of the DBI action (\ref{5dimDBI_2}) can also be written as
\begin{eqnarray}
&&\hspace{-4mm} \frac{\kappa}{2}\int d^4x dz \mbox{tr} 
\ltk K(z)^{-\frac{1}{3}} F_{\mu\nu}F_{\mu\nu}\rtk \nonumber \\
&&\hspace{-4mm}= \frac{\kappa}{2}\int d^4x dz K(z)^{-\frac{1}{3}} 
\nonumber \\
&&\hspace{-3.5mm} \times \mbox{tr}
 \biggl[ -i[\alpha_\mu,\alpha_\nu]\psi_+\psi_- +
  \lk \partial_\mu\rho_\nu -\partial_\nu \rho_\mu \rk \psi_1
  +i[\rho_\mu,\rho_\nu]\psi_1^2\nonumber\\
 &&\hspace{-3.8mm} +i\ltk 
  \lk[\alpha_\mu,\rho_\nu]\!+\![\rho_\mu,\alpha_\nu]\rk\hat{\psi}_0\psi_1
 \!+\!\lk[\beta_\mu,\rho_\nu]+[\rho_\mu,\beta_\nu]\rk\psi_1
             \rtk
	      \biggr]^2.\nonumber \\
\label{DBI_1st}
\end{eqnarray}
After diagonalizing (\ref{DBI_1st}) by using
Eqs.(\ref{eigen1})$\cdot$(\ref{normal_psi_n}),
we can finally get the effective action of pion and $\rho$ meson 
in flat four-dimensional Euclidean space-time 
from Eqs. (\ref{DBI_2nd}) and (\ref{DBI_1st}) as follows:
\begin{eqnarray}
&&\hspace{-8mm}S_{\rm D8}^{\rm DBI}-S_{\rm D8}^{\rm DBI}|_{A_M\rightarrow 0} \nonumber \\
 &&\hspace{-8mm}=\kappa\int d^4 x dz \mbox{tr}\ltk
  \frac{1}{2}K(z)^{-\frac{1}{3}}F_{\mu\nu}F_{\mu\nu}+%
  K(z) F_{\mu z}F_{\mu z}\rtk \label{DBI}\\
 &&\hspace{-8mm}=\frac{f_{\pi}^2}{4}%
 \int d^4 x \mbox{tr}\lk L_{\mu}L_{\mu}\rk%
 \hspace{22.97mm}\mbox{(chiral term)} \label{f0}\\
 &&\hspace{-7mm}+m_{\rho}^2%
 \int d^4 x \mbox{tr}\lk\rho_{\mu}\rho_{\mu}\rk%
 \hspace{21.97mm}\mbox{($\rho$-mass term)} \label{f1}\\
 &&\hspace{-7mm}+\frac{1}{2}\frac{1}{16 e^2} [\lk -i\rk^2]%
 \int d^4 x \mbox{tr}\ldk L_{\mu}, L_{\nu}\rdk^2%
 \hspace{1.87mm}\mbox{(Skyrme term)}\label{f2}\\ 
 &&\hspace{-7mm}+\frac{1}{2}%
 \int d^4 x \mbox{tr}\lk
 \partial_{\mu}\rho_{\nu}-\partial_{\nu}\rho_{\mu}\rk^2 %
 \hspace{8.2mm}\mbox{($\rho$-kinetic term)}\label{f3}\\
 &&\hspace{-7mm}+\frac{1}{2} 2 g_{3\rho} [i]%
 \int d^4 x \mbox{tr}\ltk
 \lk\partial_{\mu}\rho_{\nu}-\partial_{\nu}\rho_{\mu}\rk%
 \ldk \rho_{\mu}, \rho_{\nu}\rdk%
 \rtk%
\nonumber \\
&&
 \hspace{4.79cm}
\mbox{($3\rho$ coupling)}
\label{f4}\\
 &&\hspace{-7mm}+\frac{1}{2}g_{4\rho} [i^2]%
 \int d^4 x \mbox{tr}\ldk
 \rho_{\mu}, \rho_{\nu}\rdk^2%
 \hspace{13.8mm}\mbox{($4\rho$ coupling)}\label{f5}\\
 &&\hspace{-7mm}-i g_1%
 \int d^4 x \mbox{tr}\ltk%
 \ldk \alpha_{\mu}, \alpha_{\nu}\rdk%
 \lk\partial_{\mu}\rho_{\nu}-\partial_{\nu}\rho_{\mu}\rk%
 \rtk%
\nonumber \\
&&
 \hspace{4.243cm}
\mbox{($\partial\rho$-$2\alpha$ coupling)}\label{f6}\\
 &&\hspace{-7mm}+ g_2%
 \int d^4 x \mbox{tr}\ltk%
 \ldk \alpha_{\mu}, \alpha_{\nu}\rdk%
 \ldk \rho_{\mu}, \rho_{\nu}\rdk%
 \rtk%
 \hspace{3.2mm}\mbox{($2\rho$-$2\alpha$ coupling)}\label{f7}\\
 &&\hspace{-7mm}+ g_3%
 \int d^4 x \mbox{tr}\ltk%
 \ldk \alpha_{\mu}, \alpha_{\nu}\rdk%
 \lk%
 \ldk \beta_{\mu}, \rho_{\nu}\rdk +%
 \ldk \rho_{\mu}, \beta_{\nu}\rdk%
 \rk%
 \rtk%
\nonumber \\
&&
 \hspace{4.12cm}\mbox{($\rho$-$2\alpha$-$\beta$ coupling)} \label{f8}\\
 &&\hspace{-7mm}+i g_4%
 \int d^4 x \mbox{tr}\ltk
 \lk\partial_{\mu}\rho_{\nu}-\partial_{\nu}\rho_{\mu}\rk%
 \lk%
 \ldk \beta_{\mu}, \rho_{\nu}\rdk +%
 \ldk \rho_{\mu}, \beta_{\nu}\rdk%
 \rk%
 \rtk%
\nonumber \\
&&
 \hspace{41.5mm}
\mbox{($\rho$-$\partial\rho$-$\beta$ coupling)} \label{f9}\\
 &&\hspace{-7mm}- g_5%
 \int d^4 x \mbox{tr}\ltk%
 \ldk \rho_{\mu}, \rho_{\nu}\rdk%
 \lk%
 \ldk \beta_{\mu}, \rho_{\nu}\rdk +%
 \ldk \rho_{\mu}, \beta_{\nu}\rdk%
 \rk%
 \rtk%
\nonumber \\
&&
 \hspace{4.48cm} \mbox{($3\rho$-$\beta$ coupling)} \label{f10}\\
 &&\hspace{-7mm}-\frac{1}{2} g_6%
 \int d^4 x \mbox{tr}\lk
 \ldk \alpha_{\mu}, \rho_{\nu}\rdk +%
 \ldk \rho_{\mu}, \alpha_{\nu}\rdk%
 \rk^2%
\nonumber \\
&&
 \hspace{4.3cm}
 \mbox{($2\rho$-$2\alpha$ coupling)}\label{f11}\\ 
 &&\hspace{-7mm}-\frac{1}{2} g_7%
 \int d^4 x \mbox{tr}\lk
 \ldk \beta_{\mu}, \rho_{\nu}\rdk +%
 \ldk \rho_{\mu}, \beta_{\nu}\rdk%
 \rk^2%
\nonumber \\
&&
 \hspace{4.3cm}
 \mbox{($2\rho$-$2\beta$ coupling)}.\label{f11_1}
\end{eqnarray}
In the term (\ref{f2}), the Skyrme parameter $e$ is introduced,
comparing to the standard Skyrme model~\cite{Skyrme}, as
\begin{eqnarray}
\frac{1}{16 e^2} \!&\equiv&\! \kappa\int dz K(z)^{-\frac{1}{3}}\psi_+^2\lk 1-\psi_+\rk^2.\label{Sky_para}
\end{eqnarray}
It should be noted that the chiral term and the Skyrme term naturally
appear in terms (\ref{f0}) and (\ref{f2}) from the holographic approach.
The appearance of the chiral term is reasonable and less surprising, 
because only the chiral term is allowed as the two-derivative term of pion fields  
due to chiral symmetry and Lorentz invariance.
On the other hand, the unique appearance of 
the Skyrme term $\sim {\rm tr}[L_\mu, L_\nu]^2$ 
as the four-derivative term of pion fields 
in the holographic approach is fairly remarkable.
Actually, at the leading order of $1/N_c$ and $1/\lambda$ expansions, 
the holographic D4/D8/$\overline{\rm D8}$ QCD 
never induces other four-derivative terms 
like ${\rm tr}\{L_\mu, L_\nu\}^2$ and ${\rm tr}(\partial_\mu L_\mu)^2$ 
and higher-derivative terms
allowed by the chiral symmetry and  Lorentz invariance, 
where these terms occasionally cause 
an instability of Skyrme solitons~\cite{AF,ZB}.
The reason of the unique appearance of the Skyrme term can be naturally understood 
from the starting-point of the DBI action (\ref{5dimDBI_2}), 
because this action of $O(F^2)$ includes only `two time-derivatives' at
most, which gives a severe restriction on the possible terms 
in the effective meson theory. 
Indeed, other two candidates, 
${\rm tr}\{L_\mu, L_\nu\}^2$ and ${\rm tr}(\partial_\mu L_\mu)^2$ 
include four time-derivatives and are consequently forbidden
from the holographic point of view.
Thus, the term with four derivatives of pion fields naturally and
inevitably appears as the Skyrme term with two time-derivatives
in the holographic framework as the leading order of 1/$N_c$ and
$1/\lambda$ expansions.

In the term (\ref{f3}), owing to Eq.(\ref{normal_psi_n}), 
$\psi_1(z)$ is canonically normalized as 
\begin{eqnarray}
\kappa\int dz K(z)^{-\frac{1}{3}}\psi_1^2=1,\label{kine_normal}
\end{eqnarray}
to give the proper kinetic term of the $\rho$ meson field.
The self-couplings of $\rho$-mesons, $g_{3\rho}$ and $g_{4\rho}$,  
are expressed with the basis $\psi_1$ as 
\begin{eqnarray}
g_{3\rho}   \!&\equiv&\! \kappa\int dz K(z)^{-\frac{1}{3}}\psi_1^3, \label{g_3rho}\\
g_{4\rho}   \!&\equiv&\! \kappa\int dz K(z)^{-\frac{1}{3}}\psi_1^4. \label{g_4rho}
\end{eqnarray}
Because of the $\kappa$-dependence of the basis 
$\psi_1\propto\frac{1}{\sqrt{\kappa}}$ with
the normalization condition
(\ref{kine_normal}),
the ratio $g_{3\rho}^2/g_{4\rho}$
is 
independent of $\kappa$ 
and uniquely determined in the holographic QCD
as $g_{3\rho}^2/g_{4\rho}\simeq 0.90$, 
{\it i.e.}, $g_{3\rho}^2 \ne g_{4\rho}$.
This means that the $\rho$-meson part 
in the four-dimensional effective theory obtained from the holographic QCD 
slightly differs from the massive Yang-Mills theory,
while most phenomenological approaches adopt 
the massive Yang-Mills-type interaction ($g_{3\rho}^2= g_{4\rho}$) for the $\rho$-meson sector.

Other coupling constants $g_{1\sim 7}$ in terms (\ref{f6}$\sim$\ref{f11_1})
are defined with the basis $\psi_\pm$ (or $\hat{\psi_0}$) and $\psi_1$ as follows:
\begin{eqnarray}
g_1    \!&\equiv&\! \kappa\int dz K(z)^{-\frac{1}{3}}\psi_1\psi_+\psi_-, \label{g_1}\\
g_2    \!&\equiv&\! \kappa\int dz K(z)^{-\frac{1}{3}}\psi_1^2\lk
\frac{1}{4}-\hat{\psi}_0^2\rk, \label{g_2}\\
g_3    \!&\equiv&\! \kappa\int dz K(z)^{-\frac{1}{3}}\psi_1\psi_+\psi_- = g_1, \label{g_3}\\
g_4    \!&\equiv&\! \kappa\int dz K(z)^{-\frac{1}{3}}\psi_1^2 = 1, \label{g_4}\\
g_5    \!&\equiv&\! \kappa\int dz K(z)^{-\frac{1}{3}}\psi_1^3 = g_{3\rho}, \label{g_5}\\
g_6    \!&\equiv&\! \kappa\int dz K(z)^{-\frac{1}{3}}\psi_1^2\hat{\psi}_0^2=\frac14-g_2, \label{g_6}\\
g_7    \!&\equiv&\! \kappa\int dz K(z)^{-\frac{1}{3}}\psi_1^2 = 1. \label{g_7}
\end{eqnarray}
The holographic model has two parameters 
$\kappa=\frac{\lambda N_c}{108\pi^3}$
and Kaluza-Klein mass $M_{\rm KK}$ as the ultraviolet cutoff scale of
the theory.
$\kappa$ appears in front of the effective action (\ref{5dimDBI_2})
because the effective action of D8 brane with D4 supergravity
background expanded up to $O(F^2)$ corresponds to the leading order of 
$1/N_c$ and $1/\lambda$ expansions (see the end of subsection \ref{Non}).
Therefore, by fixing two parameters, {\it e.g.}, 
experimental inputs for $f_\pi$ and $m_\rho$, then
all the masses and the coupling constants
are uniquely determined,
which is a remarkable consequence of the holographic approach.
In particular, 
the dimensionless coupling constants (\ref{g_3rho})$\sim$(\ref{g_7}) 
are determined only by the dimensionless parameter $\kappa$.
For instance, with the typical value of $\kappa \simeq 7.460 \times 10^{-3}$, 
which reproduces the experimental ratio of $f_\pi$ and $m_\rho$ \cite{SS}, 
we numerically find 
$g_{3\rho}=g_5\simeq 5.17$,
$g_{4\rho}\simeq 29.7$,
$g_1=g_3\simeq 0.0341$, and 
$g_2=1/4-g_6\simeq 0.180$,
as well as the trivial relations $g_4=g_7=1$.

Functions $\psi_\pm(z)$ and $\psi_n(z)$ in Eq.(\ref{limit2_lr}) 
correspond to the `wave-functions' of 
pions and (axial) vector mesons in the curved fifth
dimension.
When the `oscillation' of the wave-function in the fifth dimension increases, 
it can be realized as the larger mass of the (axial) vector mesons
in the four-dimensional space-time,
which is indicated by the relation $m_n^2=\lambda_n$;
the origin of the mass is the oscillation in the extra fifth direction.
As for the pions, 
they also include slightly oscillating component $\psi_\pm$
in their wave-functions.
However the fifth coordinate is curved by D4 supergravity background,
and $\psi_\pm$ actually appear as the zero-modes in (\ref{eigen1}) with 
the curved measure.
In this sense, pions are the `geodesic' of the curved five-dimensional
space-time
and they can be realized as the massless objects.

Concerning the interaction terms with more than three bodies
of pions and (axial) vector mesons,
the wave functions of the heavier (axial) vector mesons have
smaller overlap with those of pions in the fifth direction
because of its large oscillation.
%
Such smaller overlapping of wave functions in the 
extra fifth direction between pions and `largely oscillating'
heavier (axial) vector mesons predicts
the smaller coupling constants
after the projection of the action into 
the flat four-dimensional space-time~\cite{DTSon},
which is numerically checked by the $z$-integral in the coupling
constants.
Actually, some recent experiments
indicate that the heavier (axial) vector mesons
tend to have smaller width for the decay into pions~\cite{PDG},
which is consistent with the prediction of holographic QCD 
as their smaller coupling constants
mentioned above.
Therefore, for the study of chiral solitons as the large-amplitude pion fields, 
we can expect smaller effects from 
heavier (axial) vector mesons, 
so that we consider only pions and 
$\rho$ mesons as 
the lowest massive mode ($\rho_\mu\equiv B_\mu^{(1)}$)
along the later discussions of chiral solitons in the holographic QCD.


\section{Brane-induced Skyrmions \label{Bra}}
According to the large-$N_c$ gauge theory discussed by 'tHooft~\cite{tHooft},
baryons do not directly appear as the dynamical degrees of freedom
in the large-$N_c$ limit.
In this viewpoint, there should also occur the problem of how to 
describe baryons in the large-$N_c$ holographic models of QCD.
In this section we describe a baryon as a soliton-like topological object
in the four-dimensional meson effective action (\ref{f0})$\sim$(\ref{f11_1}) induced by the 
holographic QCD,
which we now call the Brane-induced Skyrmion, 
comparing with the standard Skyrmion~\cite{Skyrme} 
based on the non-linear sigma model as its meson effective action.
Several properties of the Brane-induced Skyrmions like
meson field configurations, total energy, energy density, 
and also the size in the coordinate space
are discussed by showing the numerical results.

\subsection{Standard Skyrmions and Brane-induced Skyrmions \label{Sta}}
In 1961, Skyrme proposed the idea to describe nucleons as
mesonic solitons in a non-linear meson field theory. 
In this picture, 
a baryon is identified as a `chiral soliton' of the 
Nambu-Goldstone field,
which was called `Skyrmion', 
corresponding to the non-trivial mapping 
from the three-dimensional coordinate space to the group manifold 
$\ldk {\rm SU}(2)_L \times {\rm SU}(2)_R \rdk/{\rm SU}(2)_V\simeq {\rm SU}(2)_A$,
in accordance with spontaneous chiral-symmetry breaking.
This concept of chiral soliton provides an interesting topological picture for baryons, 
describing many properties of baryons 
with large phenomenological successes~\cite{CS}.

In terms of QCD,
which was proposed as the fundamental theory of the strong interaction
by Nambu~\cite{Nambu} in 1966 and Gross, Wilczek~\cite{GW} 
and Politzer~\cite{Poli} in 1973,
the Skyrme model had several problems;
justification for regarding the winding number
of Skyrmion as the physical baryon number, 
the ambiguity for the effective action of mesons,
and the unclear connection of it with QCD 
because the Skyrme model has actually no quarks and gluons.

In the 1970's, the analysis for large-$N_c$ QCD gives the revival of
the Skyrme model. 
In 1974, 't~Hooft prevailed that, assuming confinement,
large-$N_c$ gauge theory becomes equivalent 
to the weak coupling system of mesons (and glueballs) with
the coupling strength $g\sim 1/N_c$~\cite{tHooft}.
However, this survey gives the problem of
how baryons should appear in large-$N_c$ gauge theory.
In fact, in 1979, Witten showed that in large-$N_c$ limit 
the mass of a baryon gets increase and
proportional to $N_c$~\cite{Witten},
which explains the reason why baryons do not directly appear
as the dynamical degrees of freedom in large-$N_c$ limit. 
Witten also remarked that, in large-$N_c$ limit, 
the mass of a baryon becomes
proportional to the inverse of coupling strength of mesons;
$M_{baryon}\propto1/g$~\cite{Witten}.
As for most of the topological solitons in scalar field theories, 
their classical mass 
is proportional to the inverse of the coupling strength of
the scalar fields in the action~\cite{Raja}, which means a soliton is
a non-perturbative phenomenon on scalar fields.
These considerations 
about the large-$N_c$ gauge theory and general property of soliton waves
revived Skyrme's identification of
baryons as mesonic solitons,
and the Skyrme model was expected to be a non-perturbative (low-energy)
effective theory of QCD.

The action of the Skyrme model is originally based on
the concept of non-linear sigma model, which is
the low-energy effective theory of mesons 
reflecting spontaneous chiral-symmetry breaking 
${\rm SU}(2)_L \times {\rm SU}(2)_R \rightarrow {\rm SU}(2)_V$ in QCD.
In this model, higher mass excitations ($\rho$, $\omega$, $a_1$, {\it etc.})
are neglected
and only three physical pions, which are light pseudoscalar Nambu-Goldstone bosons 
with respect to the chiral symmetry breaking, are included as the three
independent local parameters of the coset space 
$\ldk {\rm SU}(2)_L \times {\rm SU}(2)_R \rdk/ {\rm SU}(2)_V$.
If these pion fields appear without derivatives in the effective action,
it gives the explicit breaking of the global chiral symmetry,
because constant pion fields correspond to the parameters of global chiral
transformation.
Therefore, 
pions as the local parameters of the coset space should appear accompanying
with derivative in the action, giving low-energy expansion
(or derivative expansion) for soft pions 
as the non-linear realization~\cite{Wein}.
The leading term 
respecting chiral symmetry
is uniquely defined  
containing two derivatives of pion field $U(x_\mu)$ as
\begin{eqnarray}
{\cal L}_{1}=\frac{f_{\pi}^2}{4}\mbox{tr}\lk
 \partial_{\mu}U^{\dagger}\partial^{\mu}U\rk, 
\hspace{3mm}\mbox{($U(x_\mu)\equiv e^{2i\pi(x_\mu)/f_\pi}$)}\label{S_chiral}
\end{eqnarray}
which is called the chiral term.

According to Derrick's theorem, stable soliton solutions cannot
exist only with the chiral term (\ref{S_chiral});
other terms with four or more derivatives of the pion fields are at least needed for the 
existence of stable soliton solutions in the scalar field theory in the 
three-dimensional coordinate space~\cite{Raja}.
Skyrme added a term with four derivatives as
\begin{eqnarray}
{\cal L}_{2}=\frac{1}{32 e^2}\mbox{tr}\ldk
 U^{\dagger}\partial_{\mu} U, U^{\dagger}\partial_{\nu} U\rdk^2, \label{S_skyrme}
\end{eqnarray}
which is called the Skyrme term.
Skyrme recognized the appearance of this term 
as the kinetic term of 1-form of chiral field;
$L_{\mu}=\frac{1}{i}U^{\dagger}\partial_{\mu} U$,
which can be seen by using the Maurer-Cartan equation
$\partial_\mu L_\nu-\partial_\nu L_\mu+i[L_\mu,L_\nu]=0$ 
for the Skyrme term (\ref{S_skyrme}) as
\begin{eqnarray}
{\cal L}_{2}\propto\frac{1}{2}\mbox{tr}\lk
\partial_\mu L_\nu-\partial_\nu L_\mu\rk^2.   \label{S_skyrme_kine}
\end{eqnarray}
In this viewpoint, the chiral term (\ref{S_chiral}) can be regarded as
the mass term of $L_{\mu}$~\cite{CS}.
%
Since there exist two other candidates for four-derivative terms~\cite{ZB},
the effective action up to four derivatives is not uniquely determined.

Derrick's theorem also suggests that stable soliton solutions in the scalar
field theory may exist by including other degrees of freedom 
like vector and/or spinor fields. 
Furthermore, large $N_c$ QCD 
reduces to the weak coupling system including not only 
pions but also other mesons like $\rho$, $\omega$, and $a_1$.
With these considerations, the relations of stable Skyrmions with 
vector and axial vector mesons are discussed in some context~\cite{Igarashi,MZ}.
As for the effective action between pions and (axial) vector mesons,
hidden local symmetry approach was suggested as a phenomenological
treatment for the low-energy meson dynamics~\cite{BKY}.
In this approach,  
the existence of ${\rm SU}(2)_V$ local symmetry
is assumed in the chiral Lagrangian, and $\rho$ mesons are introduced
as the gauge boson for this hidden local symmetry.
The kinetic term of $\rho$ mesons is also assumed to be dynamically generated
by some quantum effects in QCD as the dynamical pole generations.
To discuss the effect of vector mesons for the Skyrme solitons,
Igarashi {\it et al.}~\cite{Igarashi} included $\rho$ mesons 
as the dynamical gauge boson of hidden local
symmetry,
without adding four-derivative
terms like the Skyrme term, 
and showed that the kinetic term of $\rho$ mesons becomes the Skyrme term
in some parameter limit.
However there are large varieties for 
the hidden local symmetric terms, 
and action is not still uniquely determined in that framework.

Furthermore, there actually exist infinite excitation modes of 
vector and axial vector mesons. 
The consistent framework with pions and these large varieties of
(axial) vector mesons was not sufficiently developed,
and inclusive discussions about the effects of these mesons for
the Skyrme soliton properties were not achieved at all at that time.

Now, following the D4/D8/$\overline{\rm D8}$ holographic QCD discussed in the previous section,
we can uniquely get the four-dimensional meson effective action with the 
coupling terms of pions and (axial) vector mesons starting from QCD.
In this framework, massless QCD is constructed from the multi D-brane
configurations of D4/D8/$\overline{\rm D8}$ system,
and
classical supergravity description of D4 brane with D8/$\overline{\rm D8}$
probes eventually gives meson effective action 
as the low-energy effective theory of QCD.
In this effective action,
all the coupling constants are uniquely determined just by
two experimental parameters like $f_{\pi}$ and $m_{\rho}$, and 
the four-derivative term of the chiral field $U(x_\mu)$ naturally appear
as Skyrme's type~\cite{Skyrme}.

As for the (axial) vector mesons in the holographic QCD,
the hidden local symmetry,
which is embedded in the framework of Bando {\it et al.}~\cite{BKY},
naturally appears
as a part of the ${\rm U}(N_f)$ local gauge symmetry on
D8 brane with D4 supergravity background, also including the
kinetic term of the gauge field.
In analogy with `open moose model' discussed by Son {\it et al.}~\cite{DTSon},
there appear infinite excitation modes of massive (axial) vector mesons
in the holographic QCD,
through the mode expansion of the gauge field with respect to the
parity eigen-states.
Because of this mode expansion of the gauge field, 
(axial) vector mesons do not directly correspond to
the dynamical gauge boson of the hidden local symmetry, 
but they appear as some homogeneous component of it like Eq.(\ref{hidden_B}). 

In this paper we discuss the effect of $\rho$ mesons for the 
Skyrme soliton properties by using the effective action (\ref{f0})$\sim$(\ref{f11_1}),
and the contributions from the higher mass excitation modes of (axial) vector mesons
are neglected.
The reliability of this strategy is supported by the
smaller coupling constant between pions and heavier (axial) vector
mesons,
given by the oscillation of the meson wave-functions in the extra fifth
direction (see section \ref{Eff});
if we assume that the main part of the Skyrme soliton is still constructed
by the large-amplitude pion fields,
these smaller couplings may suggest the smaller effects of heavier
(axial) vector mesons for the Skyrme soliton profiles.
Although the discussion is restricted for pions and $\rho$ mesons,
the holographic QCD intrinsically includes infinite excitation modes of
(axial) vector mesons by the existence of the continuous fifth dimension.
Therefore, $\rho$ meson does not, in a strict sense,
correspond to the dynamical gauge boson of hidden local symmetry on D8
brane as discussed above.

\subsection{Hedgehog Ansatz for pion and $\rho$ meson configurations \label{Ans}}
The meson effective action (\ref{f0})$\sim$(\ref{f11_1}) induced by the holographic QCD
has ${\bf R}^3$ physical coordinate space and also the
group manifold ${\rm SU}(2)_A$ as its internal isospin space.
In order that the action (\ref{f0})$\sim$(\ref{f11_1}) is to be finite, chiral field $U({\bf x})$ and
$\rho$ meson field $\rho_{\mu}({\bf x})$ in the action should satisfy the boundary conditions as 
\begin{eqnarray}
U({\bf x})\rightarrow U_0,\hspace{4mm} \rho_{\mu}({\bf x})\rightarrow 0,
 \hspace{8mm} \mbox{(for $|{\bf x}|\rightarrow \infty$)} \label{bound1}
\end{eqnarray}
where we now conventionally take the constant matrix $U_0$ as a unit
matrix; $U_0={\bf 1}$.
The action is equal to the energy in the
Euclidean metric,
and so a configuration $U({\bf x})={\bf 1}$ and 
$\rho_{\mu}({\bf x})=0$ corresponds to a classical vacuum with
zero energy.
Therefore the boundary condition (\ref{bound1}) means that pion and $\rho$ meson fields have the same classical
vacuum for any direction at $|{\bf x}|\rightarrow \infty$;
directional variables in the coordinate space ${\bf R}^3$ become redundant at infinity.
This means that
${\bf R}^3$can be compactified at infinity
into the closed manifold $S^3$ with infinite radius. 
Now, according to the homotopical classification $\pi_3({\rm SU}(2)_A)={\bf Z}$~\cite{Steen},
the action (\ref{f0})$\sim$(\ref{f11_1}) has the Skyrme soliton solution as 
non-trivial mapping from the compactified physical space $S^3$ to
internal ${\rm SU}(2)_A$
manifold,
which is the same mechanism to the standard Skyrme model~\cite{Skyrme}.

Now we take the hedgehog Ansatz~\cite{Skyrme} for the chiral field in the
Skyrmions as
\begin{eqnarray}
U^{\star}({\bf x})=e^{i\tau_a \hat{x}_a F(r)},
\hspace{4mm}\mbox{($\hat{x}_a\equiv \frac{x_a}{r}$, $r\equiv|{\bf x}|$)}\label{HH} 
\end{eqnarray}
%
where $\tau_a$ is Pauli matrix, and $F(r)$ is a dimensionless profile function
with boundary conditions $F(0)=\pi$ and $F(\infty)=0$, giving
topological charge equal to unity. 
Ansatz (\ref{HH}) means $2\pi_a({\bf x})=\hat{x}_a F(r)$ for the pion field.
This configuration is called `maximally symmetric' in the sense that
it is invariant under a combination of space and isospace rotations~\cite{CS}.

Here we also take the hedgehog configuration Ansatz for $\rho$
meson field as
\begin{eqnarray}
&&\hspace{-4mm} \rho^{\star}_{0}({\bf x})=0, \hspace{3.2mm}
 \rho^{\star}_{i}({\bf x})=\rho^{\star}_{i a}({\bf x})\frac{\tau_a}{2}
 =\ltk \varepsilon_{i a
 b}\hat{x}_b\tilde{G}(r)\rtk
 \tau_a, \nonumber\\
&& \hspace{4.4cm}(\tilde{G}(r)\equiv
 G(r)/r)\label{WYTP}
\end{eqnarray}
where $G(r)$ is a dimensionless profile function.
This Ansatz for $\rho$ meson field is also called ``Wu-Yang-'tHooft-Polyakov
Ansatz''~\cite{Igarashi}, and
the same configuration Ansatz can be seen in the context of the gauge field 
in the 'tHooft-Polyakov monopole~\cite{Raja}.

$\rho$ meson field is sometimes described by two pion configurations as
$\rho_{ia}\sim \ldk \mbox{\boldmath
$\pi$}\times\partial_i\mbox{\boldmath $\pi$}\rdk_a$,
for example, in the hidden local symmetry approach~\cite{BKY}.
If we apply the hedgehog Ansatz (\ref{HH}) for this 
two pion configurations,
the $\rho$ meson field can be written as
\begin{eqnarray}
\rho_{ia}\sim \ldk \mbox{\boldmath $\pi$}\times\partial_i
 \mbox{\boldmath $\pi$}\rdk_a%
=\varepsilon_{abc}\pi_b\partial_i\pi_c%
=\varepsilon_{iab}\hat{x}_b \lk F^2/r\rk, \label{HLS_HH}
\end{eqnarray}
which should be compared with Eq.(\ref{WYTP}).
This comparison means that the Ansatz for pion and $\rho$ meson fields in
Eqs.(\ref{HH}) and (\ref{WYTP}) are 
consistent at least for small amplitude, from the
viewpoint of hidden local symmetry approach.
These solution ansatz as in Eqs. (\ref{HH}) and (\ref{WYTP}) are often
taken in the discussion of soliton waves~\cite{Raja};
pion and $\rho$ meson degrees of freedom are now represented by
the radial-dependent functions $F(r)$ and $\tilde{G}(r)$ respectively,
and the effective action is reduced into the one-dimensional space and
becomes solvable.

With the hedgehog Ansatz (\ref{HH}) for pion field, $L_{\mu}$, $\alpha_{\mu}$ and
$\beta_{\mu}$ appearing in the action (\ref{f0})$\sim$(\ref{f11_1}) 
can be written by using 
$\hat{\delta}_{ia}\equiv \delta_{ia}-\hat{x}_i\hat{x}_a$
and $F'(r) \equiv dF/dr$ 
as follows:
\begin{eqnarray}
&&\hspace{-4mm}L_{0}^{\star}({\bf x})=0, \label{HH_L0}\\%
&&\hspace{-4mm}L_{i}^{\star}({\bf x})=\lk \hat{x}_{i} \hat{x}_{a}F^{'}+%
  \hat{\delta}_{ia}\frac{\sin F\cdot\cos
  F}{r}+%
  \varepsilon_{iab}\hat{x}_b\frac{\sin^2F}{r}%
  \rk \tau_a, \nonumber\\
\label{HH_L}\\
&&\hspace{-4mm}\alpha_{0}^{\star}({\bf x})=0,\ 
\hspace{1.8mm}\alpha_{i}^{\star}({\bf x})=\lk\hat{x}_{i} \hat{x}_{a}F^{'}+%
                             \hat{\delta}_{ia}\frac{\sin F}{r}%
                             \rk \tau_a, \label{HH_alpha}\\
&&\hspace{-4mm}\beta_{0}^{\star}({\bf x})=0, 
\hspace{2.8mm}\beta_{i}^{\star}({\bf x})= \frac{1}{2}\varepsilon_{iab}\hat{x}_b
          \frac{1-\cos F}{r}%
	  \tau_a. \label{HH_beta}
\end{eqnarray}

\subsection{Energy and Euler-Lagrange equation for hedgehog soliton\label{Ene}}
Now, by substituting
Eqs.(\ref{HH}),(\ref{WYTP}),(\ref{HH_L0})$\sim$(\ref{HH_beta}) 
into the action (\ref{f0})$\sim$(\ref{f11_1}),
we can get the energy of the Brane-induced Skyrmion 
with the hedgehog configuration Ansatz as follows:

\begin{eqnarray}
E[F(r), \tilde{G}(r)] \!&\equiv&\! 
             \ldk S_{\rm D8}^{\rm DBI}-S_{\rm D8}^{\rm DBI}|_{A_M\rightarrow 0}
             \rdk_{{\rm hedgehog}} \nonumber \\
 &\equiv&\! \int_0^{\infty}4\pi dr r^2\varepsilon [F(r),
  \tilde{G}(r)]. \label{BIS_energy}
\end{eqnarray}
For the soliton solution, the energy $E$ is to be minimized
with the boundary conditions, $F(0)=\pi$ and $F(\infty)=0$.
The energy density multiplied by the metric $r^2$ reads
%
\begin{eqnarray}
&&\hspace{-3mm}r^2\varepsilon [F(r), \tilde{G}(r)] \nonumber \\%
&&\hspace{-3mm}=
\frac{f_\pi^2}{4}\ldk 2\lk r^2 F^{'2}+2\sin^2F \rk\rdk
\hspace{17mm}\mbox{(chiral term)}\nonumber \\
 &&\hspace{-2mm}+ m_{\rho}^2%
 \ldk 4 r^2 \tilde{G}^2\rdk
 \hspace{36.3mm}\mbox{($\rho$-mass term)}\nonumber  \\ 
 &&\hspace{-2mm}+%
 \frac{1}{32e^2}\ldk 16 \sin^2F \lk 2 F^{'2}+\frac{\sin^2F}{r^2}\rk\rdk 
 \hspace{1mm}\mbox{(Skyrme term)}\nonumber \\ 
 &&\hspace{-2mm}+\frac{1}{2}%
 \ldk 8\ltk
 3\tilde{G}^2+2 r \tilde{G}(\tilde{G}^{'}) +r^2\tilde{G}^{'2}\rtk\rdk
 \hspace{2mm}\mbox{($\rho$-kinetic term)}\nonumber \\
 &&\hspace{-2mm}- g_{3\rho}  %
 \ldk 16 r \tilde{G}^3 \rdk
 \hspace{37.41mm}\mbox{($3\rho$ coupling)}\nonumber \\
 &&\hspace{-2mm}+ \frac{1}{2}g_{4\rho}%
 \ldk 16 r^2 \tilde{G}^4 \rdk
 \hspace{33.2mm}\mbox{($4\rho$ coupling)}\nonumber \\
 &&\hspace{-2mm}+  g_1 %
 \ldk 16 \ltk F^{'}\sin F \cdot\lk
 \tilde{G}+ r \tilde{G}^{'}\rk%
 +\sin^2F\cdot\tilde{G}/r\rtk\rdk \nonumber \\
&&
 \hspace{5.132cm}\mbox{($\partial\rho$-$2\alpha$ coupling)}\nonumber \\
 &&\hspace{-2mm}- g_2%
 \ldk 16 \sin^2F\cdot\tilde{G}^2 \rdk
 \hspace{22.8mm}\mbox{($2\rho$-$2\alpha$ coupling)}\nonumber \\
 &&\hspace{-2mm}- g_3 %
 \ldk 16 \sin^2F\cdot\lk 1-\cos F\rk \tilde{G}/r \rdk 
 \hspace{2mm}\mbox{($\rho$-$2\alpha$-$\beta$ coupling)} \nonumber \\
 &&\hspace{-2mm}- g_4%
 \ldk 16 \lk 1-\cos F\rk\tilde{G}^2\rdk   
 \hspace{15.8mm}\mbox{($\rho$-$\partial\rho$-$\beta$ coupling)}\nonumber  \\
 &&\hspace{-2mm}+ g_5 %
 \ldk 16 r \lk 1-\cos F \rk \tilde{G}^3\rdk
 \hspace{17.2mm}\mbox{($3\rho$-$\beta$ coupling)}\nonumber  \\
 &&\hspace{-2mm}+ g_6%
 \ldk 16 r^2 F^{'2}\tilde{G}^2\rdk
 \hspace{26.45mm}\mbox{($2\rho$-$2\alpha$ coupling)}\nonumber  \\
 &&\hspace{-2mm}+ g_7%
 \ldk 8 \lk 1-\cos F\rk^2\tilde{G}^2\rdk
 \hspace{12.4mm}\mbox{($2\rho$-$2\beta$ coupling).}\label{energy_dense_An}
\end{eqnarray}         

Then, following the same procedure as that of
Adkins, Nappi, and Witten~\cite{ANW},
we introduce
the energy unit $E_{\rm ANW}\equiv\frac{f_{\pi}}{2e}$
and length unit $r_{\rm ANW}\equiv\frac{1}{e f_{\pi}}$,
and rewrite variables in this `Adkins-Nappi-Witten (ANW) unit' as
$\overline{E}\equiv \frac{1}{E_{\rm ANW}}E=\frac{2e}{f_\pi}E$ and
$\overline{r}\equiv \frac{1}{r_{\rm ANW}}r=e f_\pi r$.
With this scaled unit,
the energy density can be rewritten
as follows (overlines of $\overline{E}$ and $\overline{r}$ below are
abbreviated for simplicity);
\begin{eqnarray}
&&\hspace{-3mm}r^2\varepsilon [F(r), \tilde{G}(r)] \nonumber \\%
 &&\hspace{-3mm}=%
 \lk r^2 F^{'2}+2\sin^2F \rk
 \hspace{28.75mm}\mbox{(chiral term)} \nonumber \\
 &&\hspace{-2mm} + 2 \lk\frac{m_{\rho}}{f_{\pi}}\rk^2%
 \ldk 4 r^2 \tilde{G}^2\rdk
 \hspace{27.6mm}\mbox{($\rho$-mass term)}\nonumber  \\ 
 &&\hspace{-2mm} +%
 \sin^2F \lk 2 F^{'2}+\frac{\sin^2F}{r^2}\rk 
 \hspace{17.5mm}\mbox{(Skyrme term)}\nonumber \\ 
 &&\hspace{-2mm} +\lk 2e^2\rk \frac{1}{2}%
 \ldk 8\ltk
 3\tilde{G}^2+2 r \tilde{G}(\tilde{G}^{'}) +r^2\tilde{G}^{'2}\rtk\rdk
\nonumber \\
&& \hspace{5.41cm}\mbox{($\rho$-kinetic term)}\nonumber \\
 &&\hspace{-2mm} - \lk 2e^2\rk g_{3\rho}  %
 \ldk 16 r \tilde{G}^3 \rdk
 \hspace{29.2mm}\mbox{($3\rho$ coupling)}\nonumber \\
 &&\hspace{-2mm} + \lk 2e^2\rk \frac{1}{2}g_{4\rho}%
 \ldk 16 r^2 \tilde{G}^4 \rdk
 \hspace{25mm}\mbox{($4\rho$ coupling)}\nonumber \\
 &&\hspace{-2mm} + \lk 2e^2\rk  g_1 %
 \ldk 16 \ltk F^{'}\sin F \cdot\lk
 \tilde{G}+ r \tilde{G}^{'}\rk%
 +\sin^2F\cdot\tilde{G}/r\rtk\rdk \nonumber \\
&& \hspace{5.26cm}\mbox{($\partial\rho$-$2\alpha$ coupling)}\nonumber \\
 &&\hspace{-2mm} -\lk 2e^2\rk g_2%
 \ldk 16 \sin^2F\cdot\tilde{G}^2 \rdk
 \hspace{15mm}\mbox{($2\rho$-$2\alpha$ coupling)}\nonumber\\
 &&\hspace{-2mm} -\lk 2e^2\rk g_3 %
 \ldk 16 \sin^2F\cdot\lk 1-\cos F\rk \tilde{G}/r \rdk \nonumber \\
&& \hspace{5.14cm}\mbox{($\rho$-$2\alpha$-$\beta$ coupling)} \nonumber \\
 &&\hspace{-2mm} -\lk 2e^2\rk g_4%
 \ldk 16 \lk 1-\cos F\rk\tilde{G}^2\rdk   
 \hspace{7.8mm}\mbox{($\rho$-$\partial\rho$-$\beta$ coupling)} \nonumber \\
 &&\hspace{-2mm}+\lk 2e^2\rk g_5 %
 \ldk 16 r \lk 1-\cos F \rk \tilde{G}^3\rdk
 \hspace{9.4mm}\mbox{($3\rho$-$\beta$ coupling)}\nonumber  \\
 &&\hspace{-2mm}+\lk 2e^2\rk g_6%
 \ldk 16 r^2 F^{'2}\tilde{G}^2\rdk
 \hspace{18.7mm}\mbox{($2\rho$-$2\alpha$ coupling)} \nonumber \\
 &&\hspace{-2mm}+\lk 2e^2\rk g_7%
 \ldk 8 \lk 1-\cos F\rk^2\tilde{G}^2\rdk
 \hspace{4.4mm}\mbox{($2\rho$-$2\beta$ coupling).}\label{energy_dense_Re}
\end{eqnarray}         

Here we comment about the scaling property of Brane-induced
Skyrmions.
The holographic QCD with D4/D8/$\overline{\rm D8}$ system has just two
parameters like $\kappa$ and $M_{\rm KK}$.
Therefore, 
using Eqs.(\ref{f_pi_1}), (\ref{rho_mass}) and (\ref{Sky_para}),
all the parameters like 
the pion decay constant $f_\pi$,
the $\rho$ meson mass $m_\rho$ 
and the Skyrme parameter $e$
can be written by the two
parameters $\kappa$ and $M_{\rm KK}$ in the holographic framework as follows:
\begin{eqnarray}
&&\hspace{9mm}f_{\pi}=2\sqrt{\frac{\kappa}{\pi}}M_{\rm KK},\label{fpi_Mkk}\\
&&\hspace{8mm}m_\rho=\sqrt{\lambda_1}M_{\rm KK}\simeq \sqrt{0.67}M_{\rm
 KK},\label{mrho_Mkk}\\
&&\hspace{-10mm}e=\frac{1}{4}\ldk \kappa\int dz K^{-\frac{1}{3}}
                  \psi_{+}^2(1-\psi_{+})^2 \rdk ^{-\frac{1}{2}}
  \!\!\!\! \simeq\! 
  \frac{1}{4\sqrt{0.157}}\frac{1}{\sqrt{\kappa}},\label{e_kappa}
\end{eqnarray}
where the energy unit $M_{\rm KK}$ is recovered.
With these relations (\ref{fpi_Mkk})$\sim$(\ref{e_kappa}),
ANW units $E_{\rm ANW}$ and $r_{\rm ANW}$
can be written
in the holographic framework 
by the two parameters
$\kappa$ and $M_{\rm KK}$ as 
\begin{eqnarray}
E_{\rm ANW}=\frac{f_\pi}{2e}={\rm const}\cdot \kappa M_{\rm KK},\label{ANW_holo1}\\
r_{\rm ANW }=\frac{1}{ef_\pi}={\rm const}\cdot\frac{1}{M_{\rm KK}}.\label{ANW_holo2}
\end{eqnarray}
The parameter $\kappa=\frac{\lambda N_c}{108\pi^3}$
originally appears only 
in front of the effective action (\ref{5dimDBI_2})
as an overall factor of the theory because
the effective action corresponds to the leading order term of the expansion
about $1/N_c$ and $1/\lambda$ (see the end of subsection \ref{Non}).
$M_{\rm KK}$ is also the sole energy scale as the ultraviolet cutoff scale of this
holographic model.
Therefore, by taking the energy unit $E_{\rm ANW}$ $(\propto\kappa M_{\rm KK})$ as in
Eq.(\ref{ANW_holo1}),
the total energy appears as a scale invariant variable.
In fact, by introducing the rescaled $\rho$ meson field $\widehat{G}(r)$ as
\begin{eqnarray}
\widehat{G}(r)\equiv \frac{1}{\sqrt{\kappa}}\tilde{G}(r),\label{widehatG}
\end{eqnarray}
and taking into account the $\kappa$-dependence of the basis $\psi_1$
as $\psi_1\propto \frac{1}{\sqrt{\kappa}}$
indicated by the normalization condition (\ref{normal_psi_n}),
one can show that 
every energy density in each term of Eq.(\ref{energy_dense_Re})
and meson field configurations $F(r)$ and $\widehat{G}(r)$ are
scale invariant variables,
being independent of the holographic two parameters, $\kappa$ and 
$M_{\rm KK}$.

Such scaling property of Brane-induced Skyrmion can hold even by 
introducing other (axial) vector meson degrees of freedom
in this holographic framework like $a_1, \rho^{'}, a_1^{'},\cdots$
with masses $m_{a_1}, m_{\rho^{'}}, m_{a_1^{'}},\cdots$,
appearing in the mode expansion of the gauge field (\ref{limit2_lr}),
because the holographic QCD is still expressed just by 
two parameters, $\kappa$ and $M_{\rm KK}$.
Therefore, all the effects of physical parameters like
$e, f_{\pi}, m_\rho, m_{a_1}, m_{\rho^{'}}, m_{a_1^{'}},\cdots$
(or, $\kappa$ and $M_{\rm KK}$)
can be fully extracted in the units $E_{\rm ANW}$ and $r_{\rm ANW}$,
which is a remarkable consequence of the holographic approach.
Such a scaling property can also be realized in the standard Skyrme model
composed only by pion fields.
However this scaling property 
cannot be seen in the traditional phenomenological treatment
of Skyrmions with (axial) vector mesons introduced as the 
external degrees of freedom.

Euler-Lagrange equations for the pion field $F(r)$ and $\rho$ meson field
$\tilde{G}(r)$ can also be derived from 
the functional derivative of the energy $E$ with the energy density 
(\ref{energy_dense_Re}) as follows
(note that, in the ANW unit, $E$ and $r$ are dimensionless):
\begin{eqnarray}
&&\hspace{-22mm}\frac{1}{4\pi}\ltk \frac{\delta E}{\delta F(r)}-%
                   \frac{d}{dr}\lk \frac{\delta E}{\delta
                   F^{'}(r)}\rk \rtk \nonumber \\
&&\hspace{-22mm}=\lk -4rF^{'}-2r^2 F^{''}+4\sin F\cdot \cos F \rk \nonumber\\
%
&&\hspace{-22mm}+\Bigl( -4\sin F\cdot\cos F\cdot F^{'2}-4\sin^2F\cdot F^{''}\nonumber\\
&&\hspace{-22mm}+4\sin^3F\cdot \cos F\cdot 1/r^2\Bigr) \nonumber\\                
%
&&\hspace{-22mm}+ \lk 2e^2\rk g_1 \Bigl[ 16 %
               \mbox{\Large \{} 2\sin F\cdot\cos F\cdot\tilde{G}/r\nonumber\\
&&\hspace{-22mm}-\sin F\cdot \bigl( 2 \tilde{G}^{'}+r \tilde{G}^{''}\bigr)%
               \mbox{\Large \}} \Bigr] \nonumber\\
%
&&\hspace{-22mm}- \lk 2e^2\rk g_2 \ldk 16 %
               \lk 2\sin F\cdot\cos F\cdot \tilde{G}^2\rk\rdk\nonumber
\end{eqnarray}
\begin{eqnarray}
&&- \lk 2e^2\rk g_3 \Bigl[ 16 %
\bigl( \sin F + 2\sin F \cdot \cos F\nonumber\\
&&-3\sin F\cdot \cos^2F \bigr) \tilde{G}/r
                  \Bigr]\nonumber\\
%
&&- \lk 2e^2\rk g_4 \ldk 16 %
               \lk \sin F \cdot\tilde{G}^2\rk \rdk\nonumber\\
%
&&+ \lk 2e^2\rk g_5 \ldk 16 %
               \lk r \sin F \cdot\tilde{G}^3\rk \rdk\nonumber\\
%
&&+ \lk 2e^2\rk g_6 \ldk 16 %
                \lk -4r F^{'}\tilde{G}^2%
                    -2r^2 F^{''}\tilde{G}^2%
                    -4r^2 F^{'}\tilde{G}\tilde{G}^{'}\rk\rdk\nonumber\\
%
&&+ \lk 2e^2\rk g_7 \ldk 8 %
               \ltk 2 \lk 1-\cos F \rk \sin F \cdot\tilde{G}^2\rtk
               \rdk=0, \label{EL_F} 
%
\end{eqnarray}
\begin{eqnarray}
&&\frac{1}{4\pi}\ltk \frac{\delta E}{\delta \tilde{G}(r)}-%
                   \frac{d}{dr}\lk \frac{\delta E}{\delta
                   \tilde{G}^{'}(r)}\rk \rtk \nonumber \\
&&= 2\lk\frac{m_{\rho}}{f_{\pi}}\rk^2 \ldk 4 \lk 2 r^2 \tilde{G} \rk \rdk\nonumber\\
%
&&+ \lk 2e^2\rk\frac{1}{2} \ldk 8 \lk 4
\tilde{G}-4r\tilde{G}^{'}-2r^2\tilde{G}^{''}\rk \rdk\nonumber\\
%
&&- \lk 2e^2\rk g_{3\rho} \ldk 16 \lk 3r\tilde{G}^2 \rk \rdk\nonumber\\
%
&&+ \lk 2e^2\rk \frac{1}{2} g_{4\rho} \ldk 16 \lk 4 r^2 \tilde{G}^3 \rk \rdk\nonumber\\
%
&&+ \lk 2e^2\rk g_1 \ldk 16 %
\lk \sin^2 F/r-\cos F\cdot F^{'2}r-\sin F\cdot F^{''}r\rk \rdk \nonumber\\
%
&&- \lk 2e^2\rk g_2 \ldk 16 %
               \lk 2\sin^2F \cdot\tilde{G} \rk\rdk\nonumber\\
%
&&- \lk 2e^2\rk g_3 \ldk 16 %
                \sin^2 F (1-\cos F)/r \rdk \nonumber\\
%
&&- \lk 2e^2\rk g_4 \ldk 16 %
                \ltk 2(1-\cos F)\tilde{G}\rtk \rdk \nonumber\\
%
&&+ \lk 2e^2\rk g_5 \ldk 16 %
               \ltk 3r\lk 1-\cos F\rk\tilde{G}^2 \rtk \rdk\nonumber\\
%
&&+ \lk 2e^2\rk g_6 \ldk 16 %
                \ltk 2r^2 F^{'2}\tilde{G}\rtk\rdk\nonumber\\
%
&&+ \lk 2e^2\rk g_7 \ldk 8 %
               \ltk 2 \lk 1-\cos F \rk^2 \tilde{G}\rtk \rdk=0.\label{EL_G}
%
\end{eqnarray}

In $\tilde{G}(r)\rightarrow 0$ limit, Euler-Lagrange equation
(\ref{EL_F}) with the variation of pion field
coincides with that
of the standard Skyrme model without $\rho$ mesons:
\begin{eqnarray}
&&\hspace{-4.3mm}\frac{1}{4\pi}\ltk \frac{\delta E}{\delta F(r)}-%
                   \frac{d}{dr}\lk \frac{\delta E}{\delta
                   F^{'}(r)}\rk \rtk_{\tilde{G}(r)\rightarrow 0}
\nonumber \\
&&\hspace{-4mm}=2\lk -2rF^{'}-r^2 F^{''}+2\sin F\cdot \cos F \rk \nonumber\\
%
&&\hspace{-3mm}-4\lk \sin F\cdot\cos F\cdot F^{'2}+\sin^2F\cdot F^{''}
   -\frac{\sin^3F\cdot \cos F}{r^2}\rk\nonumber\\
&&\hspace{-4mm}= 0.\label{EL_F_limit}                
%
\end{eqnarray}

However even in $\tilde{G}(r)\rightarrow 0$ limit, the Euler-Lagrange
equation (\ref{EL_G}) with the variation of $\rho$ meson field
still gives a constraint for pion field $F(r)$ as follows:
\begin{eqnarray}
&&\hspace{-4.3mm}\frac{1}{4\pi}\ltk \frac{\delta E}{\delta \tilde{G}(r)}-%
 \frac{d}{dr}\lk \frac{\delta E}{\delta
 \tilde{G}^{'}(r)}\rk \rtk_{\tilde{G}(r)\rightarrow 0} \nonumber \\
&&\hspace{-4mm}= \lk 2e^2\rk g_1 \ldk 16 %
               \lk \sin^2 F/r-\cos F\cdot F^{'2}r-\sin F\cdot F^{''}r\rk \rdk\nonumber\\
&&\hspace{-3mm}- \lk 2e^2\rk g_3 \ldk 16 %
                \sin^2 F (1-\cos F)/r \rdk =0, \label{EL_G_limit} 
%
%
\end{eqnarray}
which is not included in (\ref{EL_F_limit});
Eqs.(\ref{EL_F_limit}) and (\ref{EL_G_limit}) give over-conditions for
pion field configurations.
This means that standard Skyrme configuration without $\rho$ mesons
does not correspond to 
the (local maximum, local minimum, or saddle-point) solution of
Euler-Lagrange equation (\ref{EL_F}),(\ref{EL_G}) of the 
Brane-induced Skyrme model.
In this sense,
Brane-induced Skyrme model has explicit 
instability into the non-trivial configuration of 
$\rho$ meson field ($\rho\neq 0$).
This instability is given by the correlation
processes including one $\rho$ meson, {\it i.e.},
$\partial\rho$-$2\alpha$ and
$\rho$-$2\alpha$-$\beta$ coupling terms as in Eq.(\ref{EL_G_limit}), 
affecting the dynamics of
pions even at the small amplitude limit:
$\tilde{G}(r)\rightarrow 0$.
\begin{figure}[b]
  \begin{center}
       \resizebox{65mm}{!}{\includegraphics{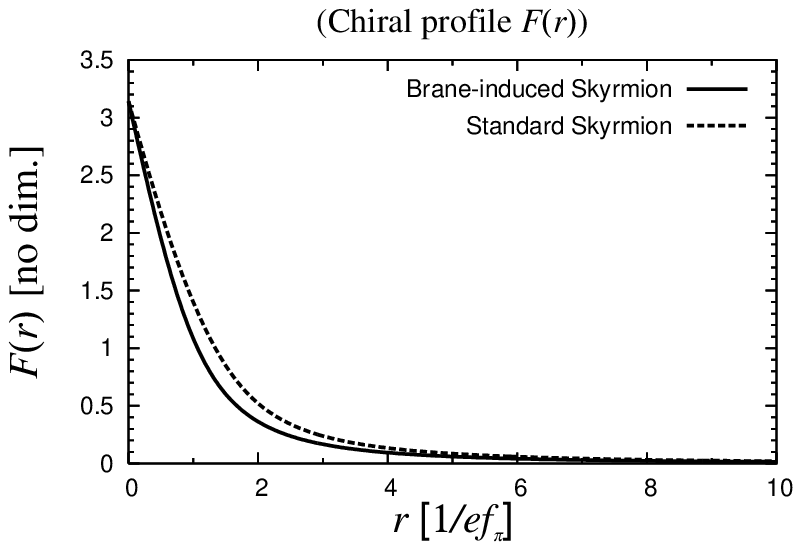}}\\
       \vspace{5mm}
       \resizebox{65mm}{!}{\includegraphics{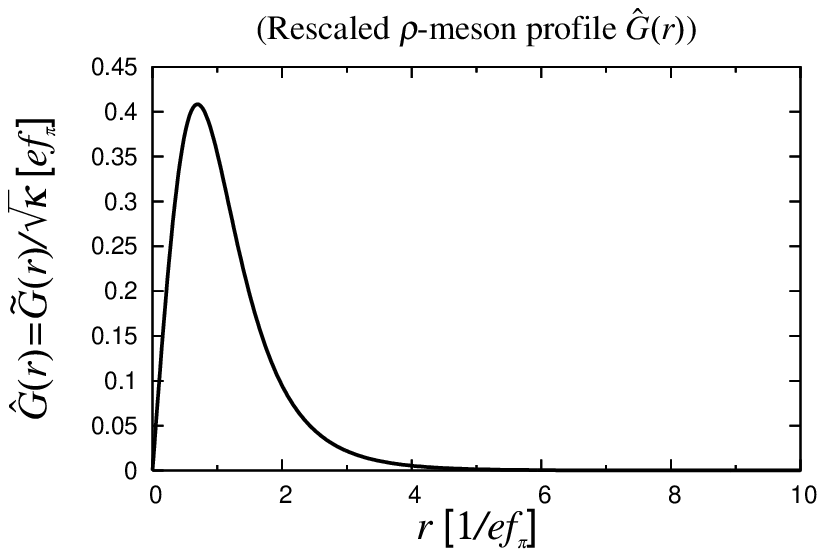}}\\
  \end{center}
\caption{ {\small Chiral profile $F(r)$ and rescaled $\rho$-meson profile
 $\widehat{G}(r)$ of Brane-induced Skyrmion as the hedgehog soliton
 solution in the holographic QCD.
 The dashed profile in the upper figure shows the chiral profile of
 standard Skyrmion without $\rho$ mesons.} }
\label{fig_conf1}
\end{figure}

\begin{figure*}[floatfix]
\begin{center}
\hspace*{-48mm}\resizebox{110mm}{!}{\includegraphics{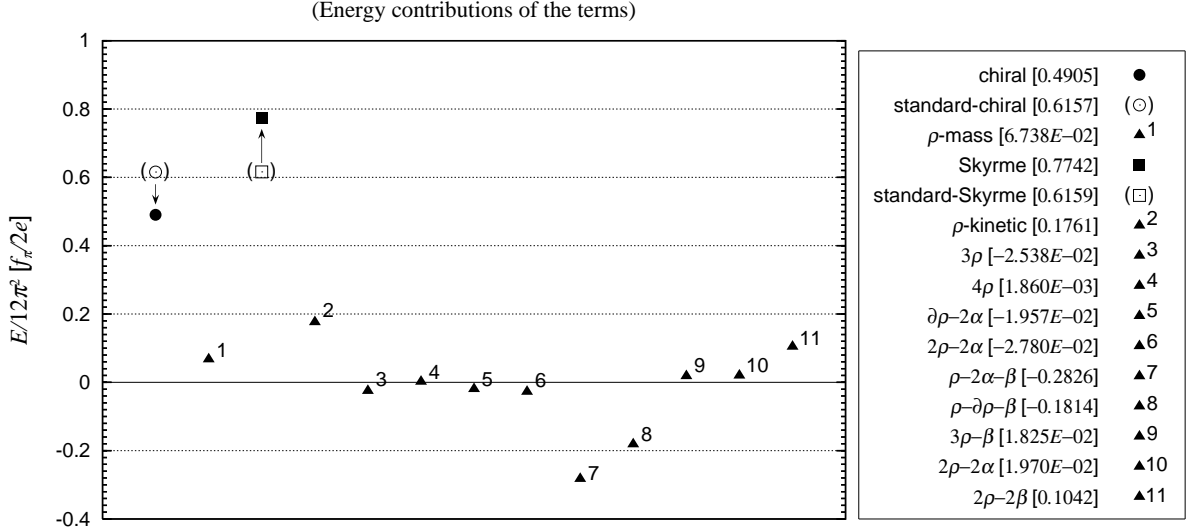}}\\
  \end{center}
\vspace{-10mm}
\caption{ {\small Energy contributions of all the terms in the action
(\ref{energy_dense_Re}) of Brane-induced Skyrmion.
Those of the standard Skyrme model with the chiral and Skyrme terms are
 also shown for comparison. } }
\label{fig_eachE1}
\end{figure*}

\subsection{Numerical results \label{Num}}
Now
we show the numerical results on the profiles of 
Brane-induced Skyrmions 
appearing in the meson effective action
(\ref{energy_dense_Re})
induced by the holographic QCD.
First we present the numerical results in ANW unit, where
meson field configurations $F(r)$ and $\widehat{G}(r)$,
total energy, energy densities, and root-mean-square radius
of Brane-induced Skyrmions are found as the scale invariant variables,
being independent of physical parameters $e$, $f_\pi$, and $m_\rho$ 
(or, $\kappa$ and $M_{\rm KK}$);
all the effects of the physical parameters are
fully extracted in the units $E_{\rm ANW}$ and $r_{\rm ANW}$
as previously discussed.
The recovering of the physical unit is discussed
in the last part of this subsection.

By numerically solving the Euler-Lagrange equations (\ref{EL_F}) and (\ref{EL_G}),
we find that the stable soliton solution exists as Brane-induced Skyrmion
with the chiral profile $F(r)$ and rescaled $\rho$-meson profile
$\widehat{G}(r)$ as in Fig.\ref{fig_conf1},
which should be a correspondent of a baryon in the large-$N_c$ holographic QCD.
To see the effect of $\rho$ meson fields
for a baryon,
the standard Skyrme configuration without $\rho$ mesons is also 
shown for comparison in Fig.\ref{fig_conf1}.
Through the interactions between pions and $\rho$ mesons in the
holographic QCD, 
pion contribution seems to be  
slightly replaced by 
$\rho$ meson degrees of freedom, 
which results in the
shrinkage of $F(r)$ from the 
standard Skyrme configuration as in Fig \ref{fig_conf1}.

The energy contributions of all the terms in the action
(\ref{energy_dense_Re}) of Brane-induced Skyrmion 
are presented in Fig.\ref{fig_eachE1}.
Those of chiral and Skyrme terms in the standard
Skyrme model 
are also shown
for comparison in this figure. 
The energy contribution of the chiral term decreases with
the shrinkage of pion field $F(r)$ 
because a trivial solution
$F(r)=0$ is most energetically favored only with the chiral term
from the Derrick's theorem~\cite{Raja}.
On the other hand, the energy contribution of the Skyrme term with four derivatives increases
with the shrinkage of $F(r)$, 
and gives the slight energy enhancement for the sum of
chiral and Skyrme terms.
This is natural because
the Skyrme configuration denoted by the dashed profile in
Fig.\ref{fig_conf1} is most stabilized in the standard Skyrme model
without $\rho$ mesons~\cite{Skyrme}.
In the Brane-induced Skyrme model, there exist the interactions
between pion and $\rho$ mesons, giving the total stability of
Brane-induced Skyrmion with non-trivial $\rho\neq 0$ configuration.


Now by solving the Euler-Lagrange equations (\ref{EL_F}) and (\ref{EL_G}),
we find the energy of Brane-induced Skyrmion in ANW unit as 
\begin{eqnarray}
E\simeq 1.115\times 12\pi^2\ldk \frac{f_{\pi}}{2e}\rdk,\label{BIS_energy}
\end{eqnarray}
which is compared with that of the standard Skyrmion
without $\rho$ mesons;
$E\simeq 1.231\times 12\pi^2 \ldk\frac{f_\pi}{2e}\rdk$,
given by Adkins {\it et al.}~\cite{ANW}. 
The total energy of Skyrmion in ANW unit is often written
as in Eq.(\ref{BIS_energy}) by the ratio relative to $12\pi^2$,
which is called the Bogomol'nyi-Prasad-Sommerfield
(BPS) saturation energy originated from the unit topological charge of
a Skyrmion (see Appendix~\ref{Ap}).
The result (\ref{BIS_energy}) suggests that the total energy of 
Brane-induced Skyrmion is reduced by $\sim 10\%$ relative to the
standard Skyrmion because of the interactions between pions and $\rho$
mesons
in the meson effective action (\ref{f0})$\sim$(\ref{f11_1}) induced by the holographic QCD.

The total energy of the standard Skyrmion is known to be slightly enhanced
by $\sim 23\%$ relative to the BPS saturation energy ($12\pi^2$)~\cite{ANW}. 
In this sense, the standard Skyrmion is called to have no BPS saturation.
The origin of this energy enhancement and non-BPS nature of the Skyrmion
was discussed from the viewpoint of 
non-linear elasticity theory by Manton {\it et al.}~\cite{Manton}.
In his survey,
the energy of a Skyrmion is regarded as a geometrical distortion energy
of the physical manifold $S_{(\infty)}^3$ with infinite radius
wrapped on the other internal manifold ${\rm SU}(2)_A\simeq S_{(\langle
\sigma\rangle=1)}^3$ with unit radius,
comparing to the `strain energy' of one deformed material winded 
on the other material. 
The two closed manifolds, 
$S_{(\infty)}^3$ and ${\rm SU}(2)_A\simeq S_{(\langle
\sigma\rangle=1)}^3$ of a Skyrmion have different radius, {\it i.e.},
different curvatures with each other.
Thus the difference of the curvatures on these two manifolds is thought
to give the
enhancement of the total energy of a Skyrmion relative to only winding energy,
{\it i.e.}, BPS saturation energy (see Appendix\ref{Ap}).
From these viewpoints, the result (\ref{BIS_energy}) may suggest that 
the metrical distortion energy of a Skyrmion is
slightly relieved by the interactions with other degrees of freedom as
$\rho$ mesons induced by the holographic QCD.
\begin{figure}[t]
  \begin{center}
   \resizebox{70mm}{!}{\includegraphics{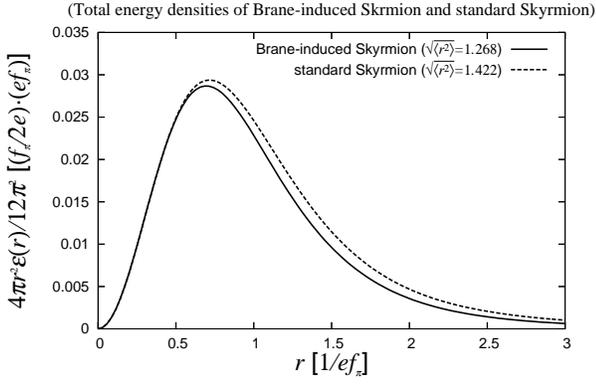}}
  \end{center}
\caption{ {\small 
Total energy densities 
$4\pi r^2 \varepsilon(r)$ 
per BPS value of $12\pi^2$ 
for Brane-induced Skyrmion (solid curve) and 
standard Skyrmion (dashed curve).
} }
\label{fig_totalEdense1}
\end{figure}
\begin{figure}[t]
   \resizebox{70mm}{!}{\includegraphics{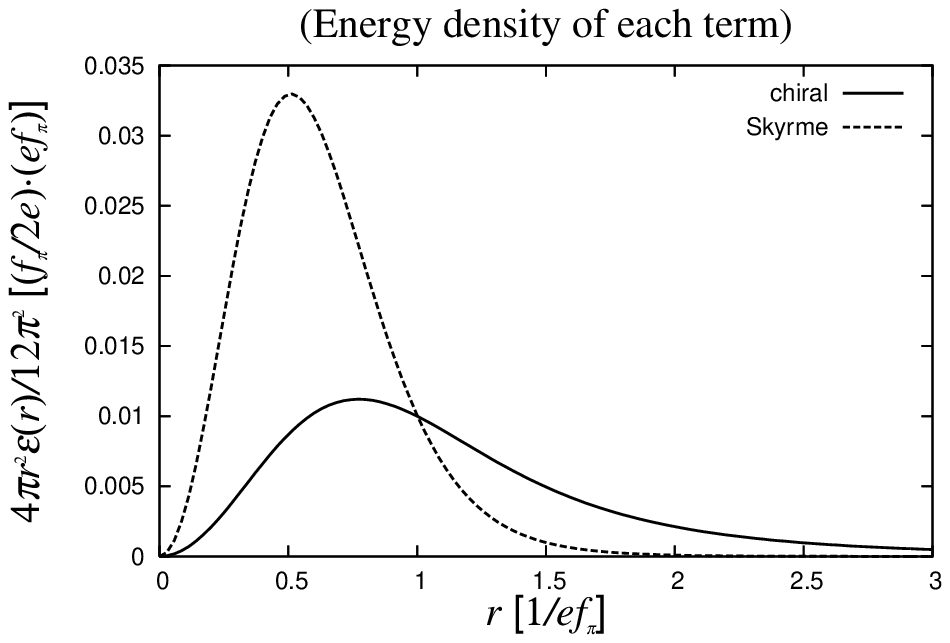}}\par
\vspace{5mm}
   \resizebox{70mm}{!}{\includegraphics{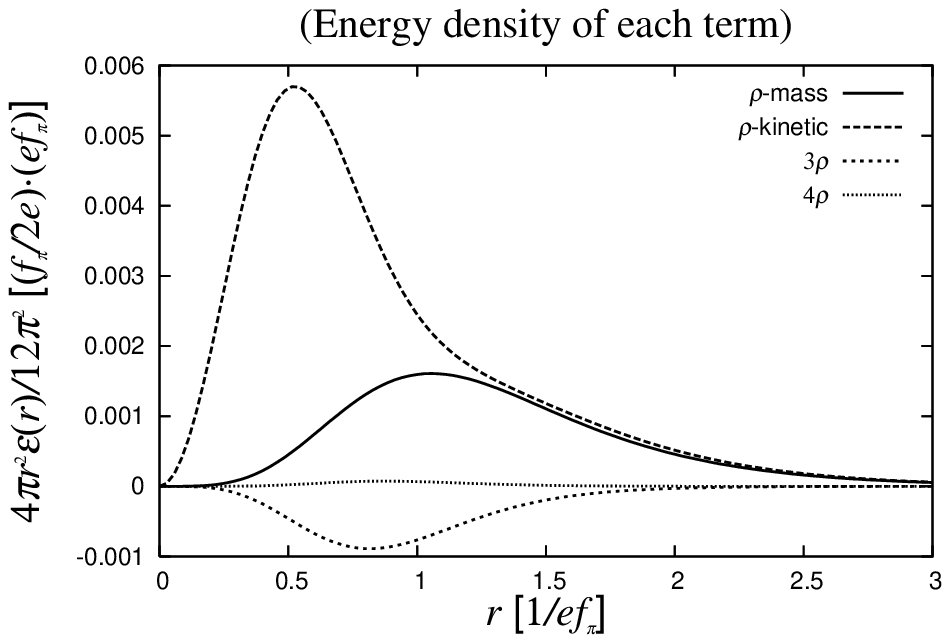}}\par
\vspace{5mm} 
   \resizebox{70mm}{!}{\includegraphics{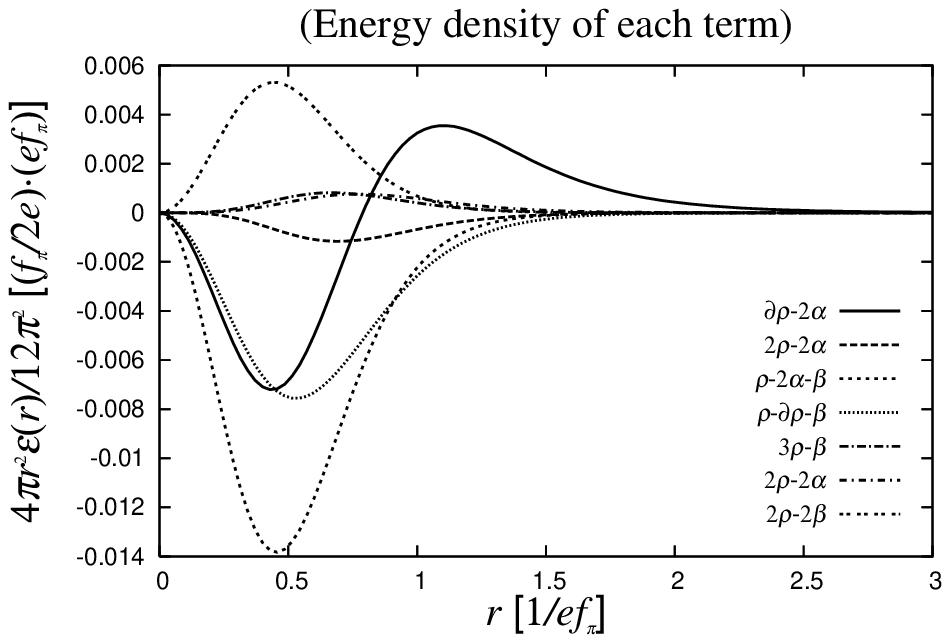}}
\caption{ {\small 
Contributions of all the terms 
in the effective action 
(\ref{energy_dense_Re}) 
to the energy density 
$4\pi r^2 \varepsilon(r)/12\pi^2$ 
of Brane-induced Skyrmion. } }
\label{fig_eachEdense1}
\end{figure}

It should also be noted that, even in $m_\rho\rightarrow \infty$ limit,
the energy of Brane-induced Skyrmion in Eq.(\ref{BIS_energy}) 
is unchanged, {\it i.e.}, it does not
coincide with that of the standard Skyrmion without $\rho$ mesons.
This $m_\rho$-independence of the total energy of Brane-induced Skyrmion
in ANW unit is originated from its scaling property 
discussed in subsection~\ref{Ene}.
In fact, $\rho$-mesons are 
naturally introduced as the internal degrees of freedom
from the fluctuation modes of open strings in the holographic framework.
Therefore only the $\rho$-meson contributions cannot be suppressed even
by taking heavy mass limit: $m_\rho\rightarrow\infty$,
in comparison with the other phenomenological treatment,
where $\rho$ mesons are included as the external degrees of freedom.

The total energy density and energy density of each term in the action
(\ref{energy_dense_Re}) of
Brane-induced Skyrmion are shown in Fig.\ref{fig_totalEdense1} and Fig.\ref{fig_eachEdense1}.
The total energy density of standard Skyrmion is also shown in
Fig.\ref{fig_totalEdense1} for comparison.
We calculate the root-mean-square radius of the Skyrme configuration
by using the normalized energy density 
$\overline{\varepsilon}(r)\equiv \varepsilon(r)/E$
($\varepsilon(r)$ is total energy density and $E$ is total energy of a
Skyrmion) as
\begin{eqnarray}
\sqrt{\langle r^2\rangle}\equiv\ldk\int_0^\infty 4\pi r^2 dr
\overline{\varepsilon}(r)r^2 \rdk^{\frac{1}{2}},
\end{eqnarray}
which gives a rough estimation of baryon size.
%
We numerically find the root-mean-square radius of Brane-induced Skyrmion in ANW unit as
\begin{eqnarray}
\sqrt{\langle r^2\rangle}\simeq 1.268 \ldk \frac{1}{ef_\pi}\rdk,\label{BIS_RMS}
\end{eqnarray}
which is compared with that of the standard Skyrmion: 
$\sqrt{\langle r^2\rangle}\simeq 1.422 \ldk \frac{1}{ef_\pi}\rdk$.
In the Brane-induced Skyrmion, 
some part of total mass is carried by the heavy vector mesons
in the soliton core,
which should give the shrinkage of the total size by $\sim 10 \%$
relative to the standard Skyrmion as in Fig.\ref{fig_totalEdense1}.

\begin{figure}[t]
 \hspace{2mm}
 \resizebox{75mm}{!}{\includegraphics{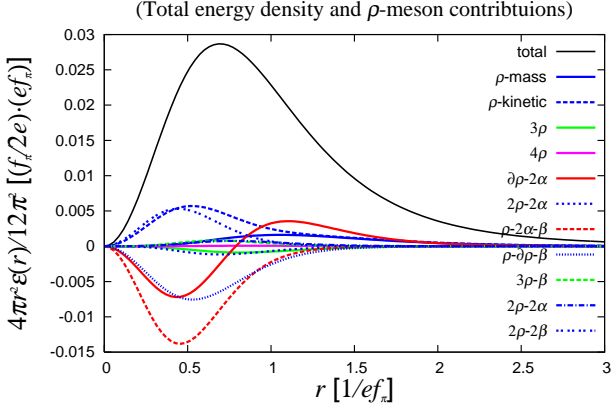}}\\
\caption{ {\small 
Contributions of 
$\rho$-meson interaction terms 
in the effective action (\ref{energy_dense_Re})
to the energy density 
$4\pi r^2 \varepsilon(r)/12\pi^2$ 
of Brane-induced Skyrmion 
with the total energy density.
} }
\label{fig_eachEdense2}
\end{figure}

The comparison between total energy density and $\rho$ meson
contributions
in the Brane-induced Skyrmion is shown in Fig.\ref{fig_eachEdense2}.
The effects of $\rho$ meson degrees of freedom for the total energy and 
root-mean-square radius of a Skyrmion are found to be relatively small with
about $10\%$ modifications discussed above.
However, Fig.\ref{fig_eachEdense2} shows that
$\rho$-meson components are rather active in the core region of
baryons (Skyrmions) 
through various interaction terms in the four-dimensional effective action.
Particularly the $\rho$-$2\alpha$-$\beta$ coupling term is found to give
about $25\%$ negative (attractive) contribution for the total baryon mass 
as shown in Fig.\ref{fig_eachE1}.
This active $\rho$-meson component inside baryons may be 
a new striking picture for baryons suggested from the holographic QCD.
Recently the projects with high-energy meson-baryon scattering are
proposed in the J-PARC experiments.
There it may be possible to see the effect of active $\rho$-meson components
in the deeper interior of baryons by using high-energy resolutions.

In this paper we do not perform the semi-classical quantization for 
the Brane-induced Skyrmions.
Nucleon $N$ and $\Delta$ isobar can be treated by isospin rotating the
Brane-induced Skyrmion as the excited hedgehog solitons.
The proper procedure for such semiclassical quantization with $\rho$ meson
fields becomes non-trivial and should be discussed in future to
get the physical interpretations of interaction channels in the action
(\ref{f0})$\sim$(\ref{f11_1}).

\begin{table*}[floatfix]
\begin{center}
\caption{\label{table_resu1} Profiles of  Brane-induced Skyrmion and standard Skyrmion with
 experimental data for baryons.}
\tabcolsep=6.6mm
\begin{tabular}[b]{cccc}\hline\hline
 & Brane-induced Skyrmion & Standard Skyrmion & Experiment \\ \hline
$f_\pi$ & $92.4$ {\rm MeV} (input) & $64.5${\rm MeV}& $92.4$ {\rm MeV}\\
$m_\rho$ & $776.0$ {\rm MeV} (input) & - & $776.0$ {\rm MeV}\\ 
$e$ & $7.32$ & $5.44$&  - \\
$E_{\rm ANW}\equiv\frac{f_\pi}{2e}$ & $6.32$ {\rm MeV} & $5.93$ {\rm MeV}& - \\
$r_{\rm ANW}\equiv\frac{1}{ef_\pi}$ & $0.29$ {\rm fm} & $0.56$ {\rm fm}& - \\
$M_{\rm HH}$ & $834.0$ {\rm MeV} & $864.3$ {\rm MeV} & - \\
$\sqrt{\langle r^2\rangle}$ & $0.37$ {\rm fm} & $0.80$ {\rm fm} &
 $0.60\sim 0.80$ {\rm fm}\\
$M_N$ & - & $938.9$ {\rm MeV} (input)&  $938.9$ {\rm MeV} \\
$M_\Delta$-$M_N$ & - & $293.1$ {\rm MeV} (input) & $293.1$ {\rm MeV} \\ \hline\hline
\end{tabular}
\end{center}
\end{table*}

Finally we discuss the recovering of the physical unit for the mass and
root-mean-square radius of Brane-induced Skyrmion.
The holographic QCD has two parameters, and now
we take $f_\pi$ and $m_\rho$ as experimental values;
\begin{eqnarray}
f_\pi=92.4 {\rm MeV},\hspace{4mm} m_\rho=776.0 {\rm MeV}\label{exp_fpi_mrho}.
\end{eqnarray}
With these two experimental inputs, 
all the variables in the holographic QCD like
$\kappa$, $M_{\rm KK}$, and $e$ are uniquely
determined by Eqs.(\ref{fpi_Mkk})$\sim$(\ref{e_kappa}) as follows:
\begin{eqnarray}
\hspace{-2mm}
\kappa\simeq 7.460\times 10^{-3},\  M_{\rm KK}\simeq 948.0
 {\rm MeV},
\ e\simeq 7.315. \label{k_Mkk_e}
\end{eqnarray}
With these variables, the static soliton energy (the baryon mass) 
and the root-mean-square radius of 
Brane-induced Skyrmion with hedgehog configuration are found to be 
\begin{eqnarray}
E \simeq 1.115\times 12\pi^2 \ldk \frac{f_{\pi}}{2e}\rdk
\!\!&\simeq&\! 834.0 {\rm MeV},\label{unit_mass}\\
\sqrt{\langle r^2\rangle}\simeq 1.268\ldk \frac{1}{ef_\pi}\rdk
\!\!&\simeq&\! 0.37 {\rm fm}.\label{unit_RMS}
\end{eqnarray}
%
With these results, Brane-induced Skyrmion in Eq.(\ref{unit_mass}) 
seems to give a reasonable baryon mass.
(The hedgehog soliton mass is slightly smaller than the nucleon mass \cite{CS}.)
However, the size in Eq.(\ref{unit_RMS}) seems to be smaller 
in comparison with 
experimental data for the proton charge radius ($\sim 0.8$fm)~\cite{exp}
and also with the size of baryon core without dynamical pion cloud
($\sim 0.6$fm) suggested by the quark model calculations~\cite{IK}.
By rotating the static hedgehog configurations of Brane-induced
Skyrmion as a semiclassical quantization procedure
and also taking into account the dynamical pion cloud,
some enhancement of the total size of Brane-induced Skyrmion is still
expected
for the argument of baryons experimentally observed like nucleons.

In the framework of the 
standard Skyrme model without $\rho$ mesons,
the masses of nucleon $N$ and $\Delta$ isobar are classically estimated
by rotating the Skyrme configuration with moment of inertia ${\cal J}$~\cite{ANW}
as
\begin{eqnarray}
M_I=M_{\rm HH}+\frac{I(I+1)}{2{\cal J}},\label{N_delta}
\end{eqnarray}
where $M_{\rm HH}$ denotes the mass of a Skyrmion as a static hedgehog
soliton, 
and isospin $I$ denotes 
$I=\frac{1}{2}$ for $N$ and $I=\frac{3}{2}$ for $\Delta$.
By taking the ANW unit, Eq.(\ref{N_delta}) can be written as
\begin{eqnarray}
M_I\ldk \frac{f_{\pi}}{2e}\rdk
=M_{\rm HH}\ldk \frac{f_{\pi}}{2e}\rdk
+(4e^4)\frac{I(I+1)}{2{\cal J}}\ldk \frac{f_{\pi}}{2e}\rdk,\label{N_delta_rescale}
\end{eqnarray}
where $e^4$ explicitly appears in the second term and, in this sense,
$N$-$\Delta$ splitting has a scale dependence even in the ANW unit.
In order to fit the $N$-$\Delta$ splitting experimentally
observed,
Adkins {\it et al.}~\cite{ANW} took a relatively small Skyrme parameter $e\simeq 5.44$,
and, to fit the experimental data for nucleon mass, 
the pion decay constant is taken as a smaller value 
$f_\pi \simeq 64.5 {\rm MeV}$.
This also gives the root-mean-square radius of a baryon 
as $\sqrt{\langle r^2\rangle}\simeq 1.422\ldk \frac{1}{ef_\pi}\rdk\simeq
0.80 {\rm fm}$.

Actually, however, $N$-$\Delta$ splitting corresponds to
the next-to-next-leading order of $1/N_c$ expansion.
Therefore the Skyrme model as the leading order $O(N_c)$ of $1/N_c$
expansion might be difficult to reproduce the experimental values
belonging to the higher-order contributions like 
$N$-$\Delta$ splitting.
This $N_c$-counting mismatch may
result in the deviation of pion decay constant $f_\pi (\simeq 64.5{\rm MeV})$
from the experimental value
by fitting the $N$-$\Delta$ splitting in the standard Skyrme model.

The comparison between the profiles of 
Brane-induced Skyrmion and
standard Skyrmion 
are summarized in Table~\ref{table_resu1} with some experimental data
for baryons.
In the holographic QCD,
the consistency on the pion decay constant $f_\pi$ and the $\rho$ meson
mass $m_\rho$ gives 
the reasonable hedgehog baryon mass and
the smaller baryon size as shown in this Table.

In the holographic QCD with D4/D8/$\overline{\rm D8}$ system,
other meson degrees of freedom like $\omega$ and $\eta^{'}$ with
${\rm U_{A}}(1)$ anomaly are discussed by introducing 
the Chern-Simons (CS) term in the effective action of D8 brane
with D4 supergravity background.
In fact,
the DBI part and CS part correspond to the same order $O(N_c)$ of $1/N_c$
expansion
because the effective action of D8 brane is composed on the D4
supergravity background as the large-$N_c$ effective theory,
and also CS part is actually the next-order contribution relative to the
DBI part with respect to the expansion about 'tHooft coupling $1/\lambda$ 
($\simeq$ 0.120 from Eqs.(21) and (110)).
%
Furthermore, most of CS part includes one time-derivative of pion fields
and does not affect the static properties of hedgehog solitons.
Nevertheless, CS term still has some time-independent part including
$\omega_0$~\cite{SS},
and it may affects soliton properties to some extent,
although CS part itself corresponds to the higher order of $1/\lambda$ expansion
as mentioned above.
Anyway, the possibility of some improvements 
in the consistency of $N$-$\Delta$ splitting and root-mean-square
radius of baryons with experimental data should be delicately discussed 
in future, even with
CS term in the holographic approach.
  
%
\section{Summary and discussion \label{Sum}}
We have studied baryons as Brane-induced Skyrmions
and numerically obtained the hedgehog soliton solution
in the four-dimensional meson effective action
induced by the holographic QCD
with D4/D8/$\overline{\rm D8}$ system.

We have reviewed the D4/D8/$\overline{\rm D8}$ holographic QCD and 
its low-energy effective meson theory 
from the viewpoints of recent hadron physics and QCD phenomenologies.
Infinite number of hidden local symmetries are shown to appear
in the holographic model
as the counterparts for the infinite tower of (axial) vector mesons,
which is consistent with the construction process of
a phenomenological
five-dimensional Yang-Mills theory from the open moose model 
by piling up infinite number of hidden local symmetries embedded in the 
chiral field~\cite{DTSon}.
Physical interpretations for the gauge fixing conditions in the 
${\rm U}(N_f)$ local gauge symmetry of D8 brane are also discussed;
$A_z=0$ gauge manifestly shows the mass generation of gauge fields to give massive (axial)
vector mesons similar to the unitary gauge in non-abelian Higgs theories,
and,
as for the residual hidden local symmetries,
$\xi_{+}^{-1}=\xi_{-}$ gauge 
gives definite parity and G-parity for the four-dimensional 
meson effective action induced by the holographic QCD. 

With the mode expansion of the gauge fields,
four-dimensional meson effective action is uniquely derived
from the holographic QCD,
without small amplitude expansion of meson fields.
We have discussed the origin of the mass of (axial) vector mesons
as the oscillations of meson wave functions in the extra fifth
dimension.
The smaller coupling constants between pions and heavier (axial) vector
mesons can also be suggested by the smaller overlapping of meson wave
functions
in the fifth dimension
because of the large oscillation of heavier (axial) vector meson wave
functions.
This is consistent with the smaller
decay width of heavier (axial) vector mesons into pions
observed in particle data group experiments. 
%
Therefore, by assuming that the main part of the chiral soliton is constructed by the large
amplitude pion fields,
the effects of higher mass excitation modes of (axial) vector mesons for 
chiral solitons are expected to be small, and so
only pions and $\rho$ mesons are treated along the discussion of
chiral solitons.

The energy functional and the Euler-Lagrange equation of Brane-induced Skyrmion
are derived with the hedgehog configuration Ansatz for
pion and $\rho$ meson fields in the four-dimensional meson effective
action induced by the holographic QCD.
The scaling property of Brane-induced Skyrmion is also discussed
by taking the ANW unit,
where all the effects of the physical parameters
are shown to be fully extracted into the units of quantities.
This scaling property can hold even by including other (axial) vector meson
degrees of freedom because
the holographic QCD has intrinsically two parameters $\kappa$
and Kaluza-Klein mass $M_{\rm KK}$ as a ultraviolet cut-off scale of the theory.
This scaling property cannot be seen in the traditional
discussion of the Skyrme model with (axial) vector mesons
included as the external degrees of freedom.

By numerically solving the Euler-Lagrange equation,
we have found that stable Skyrme soliton solution exits,
which should be a correspondent of a baryon in the large-$N_c$ holographic QCD.
Numerical results about
the meson wave functions, 
total energy, energy density, 
and root-mean-square radius of Brane-induced Skyrmion are
presented in the ANW unit
as the scale invariant properties of Brane-induced Skyrmion.
Total energy of Skyrmion tends to decrease by about $10\%$ (in ANW unit)
through the interactions of pions and $\rho$ mesons.
Total size also shrinks by about $10 \%$ (in ANW unit) with the existence
of $\rho$ mesons 
because some part of total baryon mass is carried by the 
heavy $\rho$ mesons in the core of the baryon.
By comparing the total energy density with 
$\rho$-meson contributions in the 
Brane-induced Skyrme action,
$\rho$ mesons are found to be rather active in the deeper interior of a
baryon through various interaction terms in the effective action
induced by the holographic QCD.

Finally we have recovered the physical unit for the dimensionless 
quantities by taking experimental
inputs for the pion decay constant $f_\pi$ and $\rho$ meson mass
$m_\rho$ as
$f_{\pi}=92.4 {\rm MeV}$ and $m_\rho=776.0 {\rm MeV}$.
Then the mass and root-mean-square radius of Brane-induced Skyrmion
are found to be 
$E\simeq 1.115\times 12\pi^2 \ldk\frac{f_\pi}{2e}\rdk\simeq 834.0 {\rm MeV}$
and
$\sqrt{\langle r^2 \rangle}\simeq 1.268 \ldk\frac{1}{ef_\pi}\rdk\simeq 0.37 {\rm fm}$.
Thus, the Brane-induced Skyrmion has 
a reasonable baryon mass of about 1GeV.
However, the size seems to be smaller in comparison with 
the value experimentally observed and also 
the size of the baryon core without dynamical pion cloud
suggested by quark model calculations~\cite{IK}.
By rotating the static hedgehog configuration as the semiclassical
quantization procedure, and taking into account the dynamical pion
cloud,
the enhancement in the total size of Brane-induced Skyrmion 
is still expected.

Recently there is suggested an idea 
that a baryon can be described as an instanton in the five-dimensional
Yang-Mills action in the context of the D4/D8/$\overline{\rm D8}$ holographic model~\cite{SS}.
According to the concept of Witten's baryon vertex~\cite{WittenBV},
$N_c$ fundamental strings are induced from the newly wrapped $N_c$ folded
D4 branes around the $S^4$ ($S^4$ is a four-sphere around the
extra coordinates $x_{5\sim 9}$),
which is regarded as a bound state of $N_c$ quarks as a baryon.
In the D4/D8/$\overline{\rm D8}$ holographic model, 
the existence of this newly wrapped D4 branes are represented
by an instanton on the D8 brane through the duality of 
`brane within brane'.
The complemental relations between these two kinds of topological 
non-trivialities as an instanton in
the five-dimensional space-time and Skyrme soliton in the
four-dimensional space-time derived in our framework should be 
discussed in future,
if both describe the same object as a baryon.

As for the further complicated aspects of nonperturbative QCD
with including non-zero current quark masses,
it would also be desired to construct
more simplified effective theories
from the viewpoint of the holographic QCD.
To this end,
a pioneering phenomenological construction of
the five-dimensional meson theory~\cite{DTSon,Erlich} 
may give a guide line of the framework.
It is also valuable to compare the present holographic QCD
with the analyses of QCD and SUSY-QCD
using the other type of effective theories
inspired by the AdS/CFT correspondence
\cite{Brodsky,SonStarinets}.
In any case, the holographic QCD
is one of the new powerful approaches for
the nonperturbative region of QCD,
and is expected to continuously present new striking
ideas and concepts to the hadron physics
for the future.

\section*{Acknowledgements}
Authors thank Prof. S.~Sugimoto for his inclusive lectures about holographic
QCD in Kyoto University.
Many parts of this paper are inspired during his lectures and
discussions.
Authors are also indebted to Dr. S.~Yamato for his communications about his survey
of baryons as instantons on D8 brane in the holographic model.
We are grateful to the members of Nuclear Theory Group in
Kyoto University and the Yukawa Institute for Theoretical Physics
for their discussions, supports, and encouragements.
H.S. is supported in part by the Grant for Scientific Research [(C)
No. 16540236] 
from the Ministy of Education, Culture, Sports, Science and
Technology, Japan.

\appendix
\section{Topological natures of Skyrmions\label{Ap}} 
In this Appendix, 
we summarize the topological natures of Skyrmions
from the viewpoint of manifold theories.

The total energy of standard Skyrmions is known to have a topological
lower bound,
which can be shown as the Cauchy-Schwaltz inequality relation
for the chiral and Skyrme terms in ANW unit as follows:
\begin{eqnarray}
&&\frac{1}{2}\int d^3 x \mbox{tr}(L_j L_j)-
\frac{1}{16}\int d^3 x \mbox{tr} \ldk L_j, L_k \rdk^2 \nonumber \\
&&=-\frac{1}{2}\int d^3 x \mbox{tr}(iL_j iL_j)-
\frac{1}{16}\int d^3 x \mbox{tr} \ldk iL_j, iL_k \rdk^2\nonumber\\
&&=-\int d^3 x \frac{1}{2}\mbox{tr}\ltk iL_j\pm\frac{1}{4}
                        \varepsilon_{jkl}\ldk iL_k, iL_l \rdk\rtk^2
			\nonumber \\
 &&\hspace{4mm} 
  \pm 12\pi^2\int d^3 x \frac{1}{2\pi^2}\frac{1}{2\cdot 3!}\varepsilon_{jkl}
                        \mbox{tr}(iL_j iL_k iL_l).\label{ch_Sky}
\end{eqnarray}
$L_j=\frac{1}{i}U^{\dagger} \partial_j U$ is the hermite 1-form of 
the chiral field $U\in {\rm SU}(2)_A$, which can be
expanded by the Pauli matrices as 
$L_j\equiv L_j^a\tau^a$ ($L_j^a\in {\bf R}$).
Then the first term of Eq.(\ref{ch_Sky}),
which is called a non-topological part,
is shown to be non-negative as
\begin{eqnarray}
&&(-1)^2\int d^3 x \frac{1}{2}\mbox{tr}\ltk \lk L_j^a \mp \frac{1}{4}
                        \varepsilon_{jkl}\varepsilon^{abc} L_k^b
                        L_l^c\rk\tau^a\rtk^2 \nonumber \\
&&=\int d^3 x \lk L_j^a \mp \frac{1}{4}
                        \varepsilon_{jkl}\varepsilon^{abc} L_k^b
                        L_l^c\rk^2 \geq 0. \label{non_topo1}
\end{eqnarray}

The physical meaning of the second term of
Eq.(\ref{ch_Sky}) is discussed as follows.
Skyrmion is a non-trivial mapping from the flat three-dimensional
coordinate space ${\bf R}^3$ compactified at infinity into
3-sphere $S_{(\infty)}^3$ with infinite radius,
to the group manifold ${\rm SU}(2)_A\simeq S_{(\langle \sigma
\rangle=1)}^3$
with a radius equal to the chiral condensate $\langle \sigma \rangle$,
which is conventionally taken to be a unit radius.
Regarding the second term of Eq.(\ref{ch_Sky}),
which is called a topological part,
$\frac{1}{2\cdot 3!}\varepsilon_{jkl}\mbox{tr}(iL_j iL_k iL_l)$
corresponds to the volume form on 
$S_{(\langle \sigma \rangle=1)}^3$
and
$2\pi^2$ is the total volume of $S_{(\langle \sigma \rangle=1)}^3$
with unit radius.
The ${\bf x}$ integral in this topological part runs all over the physical
coordinate space $S_{(\infty)}^3$, and therefore
topological part corresponds to $12\pi^2$ times the degree of mapping of
Skyrmions from physical space to the internal space.
Actually Skyrmion is the mapping between the two closed manifolds
$S_{(\infty)}^3$ and $S_{(\langle \sigma \rangle=1)}^3$,
and this mapping is homotopically equivalent with the integer group as
$\pi_3({\rm SU}(2)_A)={\bf Z}$~\cite{Steen}.
Therefore the degree of mapping of Skyrmions must be 
integer and, using Eq.(\ref{non_topo1}), the energy from chiral and Skyrme terms 
is found to have a topological lower bound originated from the
degree of mapping as follows:
\begin{eqnarray}
&&\hspace{-10mm} \frac{1}{2}\int d^3 x \mbox{tr}(L_j L_j)-
\frac{1}{16}\int d^3 x \mbox{tr} \ldk L_j, L_k \rdk^2
\geq 12\pi^2 |B|,\label{lower1}\\
&&\hspace{3mm} B\equiv \int d^3 x \frac{1}{2\pi^2}\frac{1}{2\cdot 3!}\varepsilon_{jkl}
                        \mbox{tr}(iL_j iL_k iL_l).\label{topo_charge1}
\end{eqnarray}
$B$ is called the topological charge and is conserved for the 
continuous distortion of waves with finite energy,
because wave should overcome an infinite energy barrier
to cross into the different homotopical sectors.
$B$ is identified as the conserved baryon number in the Skyrme model~\cite{Skyrme},
which prevents a proton from decaying into pions~\cite{Manton}.
Therefore standard Skyrmion with a unit degree, which corresponds to one
baryon
with $B=1$,
has topological lower bound for its energy as $E\geq 12\pi^2$.

Regarding several solitons (Skyrmions, monopoles, instantons, {\it etc.}),
they are realized as some kinds of topological
non-triviality,
and their energies tend to have a lower bound originated from its
topological charge.
If the soliton energy is exactly equal to this topological lower bound,
then this soliton is called to have Bogomol'nyi-Prasad-Sommerfield
(BPS) saturation~\cite{Raja}.
BPS saturation is known to occur in some cases of solitons with the
gauge theories like 'tHooft-Polyakov monopole in the ${\rm SU}(2)$ Higgs
model,
and also the instantons with the Euclidean metric.

As for the case of Skyrmions,
Adkins et.al.~\cite{ANW} numerically showed that 
standard Skyrmion has no BPS saturation, and
there exists slight enhancement of total energy by $\sim 23\%$
relative to the BPS saturation energy;
$E\simeq 1.231\times 12\pi^2 \ldk\frac{f_\pi}{2e}\rdk$. 
The origin for this enhancement of total energy and non-BPS nature of
Skyrmions is discussed from the viewpoint of 
non-linear elasticity theory by Manton {\it et al.}~\cite{Manton}.
In this survey, the energy of 
the Skyrmion is regarded as a measure of the geometrical distortion
induced by the mapping between
two manifolds $S_{(\infty)}^3$ and $S_{(\langle \sigma \rangle=1)}^3$,
which is compared to the strain energy of one deformed material
winded on the other material.
If the two manifolds are not isometric with each other,
the distortion energy of one manifold winded around the other one
is enhanced by the difference of curvatures,
relative to the only winding energy, {\it i.e.},
BPS saturation energy.
Standard Skyrmion is a mapping between two non-isometric manifolds;
$S_{(\infty)}^3$ with infinite radius and $S_{(\langle \sigma \rangle=1)}^3$
with unit radius,
which then gives the enhancement of Skyrmion energy relative to the
BPS saturation energy.

By solving the Euler-Lagrange equations (\ref{EL_F}) and (\ref{EL_G})
induced by the holographic model,
we find the energy of Brane-induced Skyrmion as 
\begin{eqnarray}
E\simeq 1.115\times 12\pi^2\ldk \frac{f_{\pi}}{2e}\rdk,\label{BIS_energy_ap}
\end{eqnarray}
which is lower by $\sim 10\%$ relative to that of standard Skyrmion.
This may be understood that the metrical distortion energy is
slightly relieved by the interactions with other degrees of freedom as
$\rho$ mesons,
which is discussed in Eq.(\ref{BIS_energy}).


\end{document}